\documentclass[twocolumn]{IEEEtran}
\usepackage{amsmath}
\usepackage{amssymb}
\usepackage{amsfonts}
\usepackage{color}
\usepackage{graphicx}
\usepackage{subcaption}
\usepackage{mathrsfs}
\usepackage{times}
\usepackage{empheq}
\usepackage{epstopdf}
\usepackage{cite}
\usepackage{relsize}
\usepackage{relsize}
\usepackage{arydshln}
\usepackage{float}
\usepackage[normalem]{ulem}
\usepackage{mathtools,amssymb,lipsum}
\usepackage{cuted}
\usepackage{lettrine}
\newcounter{MYtempeqncnt}
\title{Complex-Valued Symbol Transmissions in Filter Bank Multicarrier Systems using Filter Deconvolution}
\author{Adnan Zafar, Mahmoud Abdullahi, Lei Zhang, Sohail Taheri, Pei Xiao and Muhammad Ali Imran
	\thanks{A.Zafar, M.Abdullahi, S.Taheri and P.Xiao are with Institute for Communication Systems (ICS), University of Surrey, Guildford, UK. Emails: \{a.zafar, m.abdullahi, s.taheri, p.xiao\}@surrey.ac.uk. A. Zafar is also affiliated with Institute of Space Technology, Islamabad, Pakistan.}
	\thanks{L.Zhang and M.A.Imran are with School of Engineering, University of Glasgow, Glasgow, UK. Email: \{lei.zhang, muhammad.imran\}@glasgow.ac.uk}
}

\begin{document}
\maketitle
\begin{abstract} Transmission of complex-valued symbols using filter bank multicarrier systems has been an issue due to the self-interference between the transmitted symbols both in the time and frequency domain (so-called intrinsic interference). In this paper, we propose a novel low-complexity interference-free filter bank multicarrier system with QAM modulation (FBMC/QAM) using filter deconvolution. The proposed method is based on inversion of the prototype filters which completely removes the intrinsic interference at the receiver and allows the use of complex-valued signaling. The interference terms in FBMC/QAM with and without the proposed system are analyzed and compared in terms of mean square error (MSE). It is shown with theoretical and simulation results that the proposed method cancels the intrinsic interference and improves the output signal to interference plus noise ratio (SINR) at the expense of slight enhancement of residual interferences caused by multipath channel. The complexity of the proposed system is also analyzed along with performance evaluation in an asynchronous multi-service scenario. It is shown that the proposed FBMC/QAM system with filter deconvolution outperforms the conventional OFDM system.
\end{abstract}
%SECTION I-INTRODUCTION
\section{Introduction}
\lettrine{I}{}ncreasing demands for higher data rates in mobile communication and 5G application requirements such as Internet of Things (IoT), Gigabit wireless connectivity, and tactile internet present an ultimate challenge to provide a uniform service experience to users \cite{5GNow2014Wunder, LeiMultiservice2017}. To this end, the new physical layer should provide two important features. First, variably aggregation of non-adjacent bands to acquire higher bandwidths for data transmission \cite{ CABoguka2015}. Second, supporting asynchronous transmissions, reducing signaling overhead and handling sporadic traffic generating devices such as IoT devices \cite{FSMattera2015}. The features necessitate a new waveform which provides very low out of band radiation (OoBR), as well as immunity against synchronization errors.\\
\indent As orthogonal frequency division multiplexing (OFDM) is unable to satisfy the new physical layer requirements, several waveforms have been introduced as a potential replacement for it. Filter bank multicarrier with offset quadrature amplitude modulation (FBMC/OQAM) is one of the promising candidates which provide very low OoBR, as well as immunity against synchronization errors, thanks to its per-subcarrier filtering \cite{5gzte}. The main drawback in FBMC/OQAM is that it relaxes the orthogonality condition to \emph{real} field to utilize a well-localized filter in time and frequency, and maintain transmission at the Nyquist rate \cite{FBMC1}. This is because according to Balian-Low theorem \cite{ BalianDaubechies90, BalianBenedetto95,FBMC1}, there is no way to utilize a well-localized prototype filter in both time and frequency, along with maintaining orthogonality and transmitting at Nyquist rate. Thus, relaxing the orthogonality condition (OQAM modulation) can guarantee the other two factors. Consequently, the transmitted real symbols in this system are contaminated with imaginary interference terms (intrinsic interference) at the receiver. The intrinsic interference is the main issue for FBMC/OQAM transceivers. First of all, in highly dispersive channels, the system will not perform properly with single-tap equalization \cite{FBMCLei}. Secondly, multiple-input-multiple-output (MIMO) applications such as maximum likelihood detection \cite{EQ2}, and the Alamouti space-time block coding \cite{SD1} are not directly applicable to the system. Finally, due to intrinsic interference, channel estimation process in FBMC/OQAM is not as straightforward as OFDM systems. To facilitate channel estimation, it is necessary for the transceiver to perform further pilot processing or waste part of the transmit resources \cite{taheriCFC, taherizte }.\\  
\indent The idea behind FBMC with QAM modulation is to reach a quasi-orthogonal signal while maintaining per-subcarrier filtering. There are two types of this system in the literature. Type I which was introduced in \cite{qam-fbmc-nam-ICC2014}, uses two different prototype filters for odd and even subcarriers to mitigate intrinsic interference. One of the proposed filters in \cite{qam-fbmc-nam-ICC2014} suffers from very poor OoBR which was enhanced in \cite{qam-fbmc-yun-spawc2015} and \cite{qam-fbmc-kim-globecom2015}. Type II of FBMC/QAM introduced in \cite{qam-fbmc-yun-spawc2015} uses an optimized prototype filter for all subcarriers. The advantage of Type II is that the OoBR rapidly decays to the desired level within one subcarrier spacing, which is an imposed constraint on the cut-off frequency of the prototype filter stated in \cite{ FBMCphydBellang2001}. This method is also known as filter bank based OFDM (FB-OFDM) in \cite{fb-ofdm}. Nevertheless, the filter design in this type of system is quite critical in order to achieve an acceptable level of orthogonality, while keeping the Nyquist property in the time domain. \\ 
\indent In this paper, we target at the filter design for type II of FBMC/QAM systems due to its mentioned advantage. To mitigate the energy of intrinsic interference in this system, a remedial system is required which is known as \emph{inverse system} in the general context of linear systems theory \cite{ Proakis:DSP }. The inverse system, is cascaded with the multicarrier filtering, and thus yields a replica of the transmitted symbols without interference terms, after channel equalization. Since the inverse system counteracts the effect of multicarrier filtering, the process is called \emph{deconvolution}. In this process, the transmitted symbols are separated from the filtering characteristics of the system. 
We propose this novel interference-free FBMC system based on inversion of the prototype filters. The advantage of this system is that it can retain the positive features of FBMC and OFDM at the same time e.g. the channel estimation and equalization can be performed in a straightforward way as in OFDM together with other advantages that can be achieved in FBMC systems, such as low OoBR and robustness to synchronization errors. \\%while low OoBR and robustness to synchronization errors can be achieved as in FBMC.\\
The main contributions in this work can be itemized as follows
\begin{itemize}
\item A matrix model of the QAM based FBMC system is presented in the presence of additive noise and multipath channel. The interference terms at the receiver due to channel distortions and the intrinsic behavior of the transceiver model are also derived.
\item An inverse filter matrix based on prototype filters is then introduced at the receiver to cancel the effects of intrinsic interference in the FBMC/QAM system. It has been shown with theoretical analysis that the introduction of inverse filter completely removes the intrinsic interference.
\item The interference terms including the ones introduced by the multipath channel are analyzed in terms of mean square error (MSE) with and without the inverse filter. It is also shown that the interference cancellation process significantly improves the system output signal to interference plus noise ratio (SINR). %has negligible effect on the noise and interference enhancement in the system.
\item Complexity analysis of the FBMC/QAM system with and without the inverse filter is also presented. It is shown that the receiver complexity in both cases have the same upper bounds.
\end{itemize}

The rest of this paper is organized as follows. The system model of the FBMC/QAM system with and without interference cancellation, as well as the derivation of the interference terms are provided in Sec. II. The interference and complexity analysis %with and without the inverse system at the receiver 
are presented in Sec. III and IV respectively. In Sec. V, the proposed system is then evaluated for an asynchronous multi-service scenario and the performance is compared with conventional OFDM system. Finally, the conclusions are drawn in Sec. VI.
\paragraph*{Notations} Vectors and matrices are denoted by lowercase and uppercase  bold letters. $\{\cdot\}^H, \{\cdot\}^T, \{\cdot\}^*$ stand for the Hermitian conjugate, transpose and conjugate operation, respectively. $\mathcal {E}\{\mathbf A\}$ denotes the expectation operation of $\mathbf A$. $\mathcal {F}$ and $\mathcal F^H$ represents the power normalized N point discrete Fourier transform (DFT) and inverse DFT (IDFT) matrices. $\mathbf I_{m\times m}$ refers to $m$ dimension identity matrix and for some cases the subscript will be dropped for simplification whenever no ambiguity arises. $\|\mathbf A\|_n^2$ means taking the $n^{th}$ diagonal element of matrix $\|\mathbf A\|^2=\mathbf A \mathbf A^H$. %$A\otimes B$ represents kronecker product of $A$ and $B$. 
We use $*$ as a linear convolution operator. $\textrm{Tr}\{\mathbf A\}$ denotes the trace of matrix $\mathbf A$.
%SECTION II-FBMC/QAM SYSTEM
\section{FBMC/QAM System}\label{sec2}
In this section we define the FBMC/QAM system in matrix form which will be subsequently used to propose an inverse system based on prototype filters to cancel the effect of intrinsic interference. %The complete system model along with inverse filter is presented in the sequel.
%SECTION II-A SYSTEM MODEL
\subsection{System Model}
The system model is divided into transmit processing, multipath channel and receive processing blocks as follows
%SECTION II-A-I Transmit Processing
\subsubsection{{Transmit Processing}}
\begin{figure}[th]
	\centering
	\includegraphics[scale=0.5]{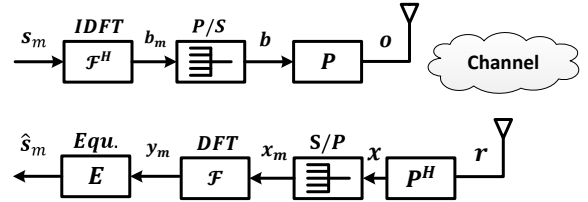}
	\caption{Block diagram of FBMC/QAM system}
	\label{fig1}
\end{figure}

The FBMC/QAM system follows a block based processing approach where each block contains $M$ FBMC/QAM symbols and each symbol has $N$ subcarriers in the frequency domain i.e. each block is represented as ${S}=[{\mathbf{s}_0},{\mathbf{s}_1},{\mathbf{\cdots}},{\mathbf{s}_{M-1}}]\in \mathbb{C}^{N\times M}$ where ${\mathbf{s}_m}=[{s_{m,0}},{s_{m,1}},{\cdots},{s_{m,N-1}}]^T\in \mathbb{C}^{N\times1}$. %The transmitted signal on the $n^{th}$ subcarrier in a FBMC/QAM system is a scalar. 
Hence, the total number of QAM symbols transmitted in one FBMC/QAM block is $MN$. Furthermore, the power of the modulated symbol $s_{m,n}$ is represented as $\delta^2$ i.e., $\mathcal{E}\{\|s_{m,n}\|^2\}= \delta^2$. The block diagram for both transmitter and receiver of FBMC/QAM is shown in Fig {\ref{fig1}}.
According to  Fig. \ref{fig1}, the signal $\mathbf{{s}}_m$ is passed through an $N$ point IDFT processor and can be expressed as 
\begin{eqnarray}\label{eq:1}
{\mathbf{{b}}}&=&[{\mathbf{{b}}_0};{\mathbf{{b}}_1};{\mathbf{\cdots}};{\mathbf{{b}}_{M-1}}]\nonumber \\
&=&[{\mathbf{\mathcal{F}}^H}{\mathbf{{s}}_0};{\mathbf{\mathcal{F}}^H}{\mathbf{{s}}_1};{\mathbf{\cdots}};{\mathbf{\mathcal{F}}^H}{\mathbf{{s}}_{M-1}}]\in \mathbb{C}^{MN\times1}.
\end{eqnarray}
%SECTION II-A-II Prototype Filters / Filter Matrices
\subsubsection{{Prototype Filter / Filter Matrix}}
The signal then passes through a prototype filter $\mathbf{w}$. It has been reported that a well-designed prototype filter with moderate length (e.g., overlapping factor $K$ = 4 $\sim$ 6) incurs negligible %self-
interference \cite{6849669}. To generalize our derivation, let us suppose the filter overlapping factor is $K$, so the total length of the prototype filter is $KN$ i.e., ${\mathbf{w}}=[{\mathbf{w}_0},{\mathbf{w}_1},\cdots,{\mathbf{w}_{K-1}}]=[{w}_0,{w}_0,\cdots,{w}_{KN-1}]\in \mathbb{R}^{1\times KN}$. In general, the prototype filters are linearly convolved with the input signal but to represent the complete system in matrix form we have to present the filtering process in matrix form as well. % such that 
The multiplication of the filter matrix with the input vector is equivalent to the required linear convolution process. The prototype filter matrix $\mathbf {P} \in \mathbb{R}^{(K+M-1)N \times MN}$ is therefore defined as
\begin{eqnarray}\label{eq:2}
\mathbf {P} =
\begin{pmatrix}
\mathbf {W}_0 & \mathbf {0}& \mathbf {0}& \cdots & \mathbf {0} \\
\mathbf {W}_1 & \mathbf{W}_0& \mathbf {0}& \cdots & \mathbf {0}  \\
\vdots & \mathbf{W}_1& \mathbf{W}_0& \cdots & \mathbf{0}\\
\mathbf {W}_{K-1} & \vdots & \mathbf{W}_1&  \cdots & \mathbf{0}\\
\mathbf {0} & \mathbf {W}_{K-1} & \vdots &  \cdots & \mathbf {W}_0\\
\mathbf {0} & \mathbf {0}& \mathbf {W}_{K-1}  &  \cdots & \mathbf {W}_1\\
\vdots & \vdots& \vdots  &  \ddots & \vdots\\
\mathbf {0} & \mathbf {0}& \mathbf {0}  &  \cdots & \mathbf {W}_{K-1}\\
\end{pmatrix},
\end{eqnarray}
where $\mathbf{W}_k=diag(\mathbf{{w}}_k)\in \mathbb{R}^{N\times N}$ for $k=0,1,2,\cdots,K-1$ and $\mathbf{{w}}_k=[{w}_{kN},{w}_{kN+1},\cdots,{w}_{kN+N-1}]\in \mathbb{R}^{1\times N}$. The output of the filter matrix $\mathbf {P}$ is formed as
\begin{eqnarray}\label{eq:3}
{\mathbf{o}}={\mathbf{P}{\mathbf{b}}}\in \mathbb{C}^{(K+M-1)\times1}.
\end{eqnarray}
The output of the filter i.e., $\mathbf{o}$ has $(K-1)N$ more samples due to the linear convolution process.
%SECTION II-A-III Channel Impulse Response
\subsubsection{{Channel Impulse Response}}
We assume the system operates over a slowly-varying fading channel i.e., quasi-static fading channel. In such a scenario, we can assume that the duration of each of the transmitted data block is smaller than the coherence time of the channel, therefore the random fading coefficients stay constant over the duration of each block \cite{tse2005fundamentals}. In this case, we define the multipath channel as a $L$-tap channel impulse response (CIR) with the $l^{th}$-tap power being ${\rho^2_l}$. It is also assumed that the average power remains constant during the transmission of the whole block. Let us define the CIR $\mathbf h$ as
\begin{eqnarray}\label{eq:4}
\mathbf h = [h_{0},h_{1},\cdots,h_{L-1}]^T= [\rho_0 z_{0},\rho_1 z_{1},\cdots,\rho_{L-1} z_{L-1}]^T,
\end{eqnarray}
where $h_{l}$ defines the $l^{th}$ tap in the time domain CIR and the complex random variable $z_l$ with complex Gaussian distribution as $\mathbb{C}\mathcal{N}(0,1)$ represents the multipath fading factor of the $l^{th}$ tap of the quasi-static rayleigh fading channel. 
%SECTION II-A-IV Passing through the channel
\subsubsection{{Passing through the Channel}}
The signal $\mathbf{{o}}$ after the prototype filtering is then passed through the channel $\mathbf h$. The received signal is now represented as
\begin{eqnarray}\label{eq:5}
\mathbf  {r} &=& \mathbf h * \mathbf {o} + \mathbf n,
\end{eqnarray}
where $\mathbf{n}$ is Gaussian noise with each element having zero mean and variance $\sigma^2$. To represent the convolution process given in (\ref{eq:5}) as matrix multiplication, we first define the $l^{th}$ tap multipath fading factor $\mathbf z_l$ in a diagonal matrix form as follows:
\begin{eqnarray}\label{eq:6}
\mathbf{Z}_{l}= z_{l}\times \mathbf I_{(K+M-1)N \times (K+M-1)N}. %\in\mathbb{C}^{(K+M-1)N\times (K+M-1)N}
\end{eqnarray}
The definition of $\mathbf{Z}_{l}$ implies that each FBMC/QAM symbol in a block experiences the same channel. With all these definitions we can reform (\ref{eq:5}) as
\begin{eqnarray}\label{eq:7}
\mathbf  {r} =\sum_{l=0}^{L-1} \rho_l \mathbf Z_{l}\mathbf {o}^{\downarrow l} + \mathbf o_{IBI}+\mathbf n,
\end{eqnarray}
where $\mathbf o_{IBI}=\sum_{l=0}^{L-1}\rho_l \mathbf Z_l\mathbf r_{B,l}$ is the inter-block interference (IBI) caused by  channel multipath effect with $\mathbf r_{B,l} = [\mathbf r_{p,l}; \mathbf 0_{[(M+K-l)N-l]\times 1}]$ and $\mathbf r_{p,l} \in \mathbb{C}^{l\times 1}$ is the interfering signal from the previous FBMC/QAM block. When guard time is longer than the channel duration, we have $\mathbf r_{B,l} = \mathbf 0$ and consequently, $\mathbf o_{IBI} = \mathbf 0$. In (\ref{eq:7}), $\mathbf {o}^{\downarrow l}$ represents $l$-sample delayed version of $\mathbf{{o}}$ with zero padding in the front and is represented as $\mathbf {o}^{\downarrow l}=[\mathbf 0_{l\times 1};\mathbf{{o}}_{q,l}]$. Where $\mathbf{{o}}_{q,l}$ represents the first $(K+M-1)N-l$ elements of $\mathbf{{o}}$.  From (\ref{eq:3}) we can write $\mathbf {o}^{\downarrow l}=\mathbf {P}^{\downarrow l}\mathbf{{b}}$, where $\mathbf {P}^{\downarrow l}=[\mathbf 0_{l\times MN}; \mathbf{P}_{q,l}]$. Here $\mathbf{P}_{q,l}$ is the first $(K+M-1)N-l$ rows of $\mathbf {P}$. We can thus reform (\ref{eq:7}) as follows:
\begin{eqnarray}\label{eq:8}
\mathbf  {r} = \sum_{l=0}^{L-1} \rho_l \mathbf Z_l\mathbf{P}^{\downarrow l} \mathbf{b} + \mathbf o_{IBI} + \mathbf n.
\end{eqnarray}
Eq (\ref{eq:8}) indicates that as a result of channel multipath effect, the original $\mathbf{P}$ is replaced by distorted filter matrix $\mathbf{P}^{\downarrow l}$. In order to demonstrate the relationship of the distortion and the multipath effect on the FBMC/QAM system, we first introduce a block diagonal exchanging matrix $\mathbf X_l \in \mathbb{R}^{MN\times MN}$ as follows:
\begin{eqnarray}\label{eq:9}
\mathbf {X}_{l} =
\begin{bmatrix}
\mathbf {X}_{sub,l} & \mathbf {0} & \cdots & \mathbf {0} \\
\mathbf {0} & \mathbf {X}_{sub,l} & \cdots & \mathbf {0} \\
\vdots  & \vdots  & \ddots & \vdots  \\
\mathbf {0} & \mathbf {0} & \cdots & \mathbf {X}_{sub,l}
\end{bmatrix}
\in\mathbb{R}^{MN\times MN},
\end{eqnarray}
with
\begin{eqnarray}\label{eq:10}
\mathbf {X}_{sub,l} =
\begin{bmatrix}
\mathbf {0}_{l\times (N-l)} & \mathbf {I}_{l\times l}  \\
\mathbf {I}_{(N-l)\times (N-l)} & \mathbf {0}_{(N-l)\times l}
\end{bmatrix}
\in\mathbb{R}^{N\times N}.
\end{eqnarray}
As $\mathbf {X}^T_{l}\mathbf {X}_{l} = \mathbf I$, we have
\begin{eqnarray}\label{eq:11}
\mathbf{o}^{\downarrow l} = \mathbf {{P}}^{\downarrow l}\mathbf{b}=\mathbf {{P}}^{\downarrow l}\mathbf {X}^T_{l}\mathbf {X}_{l} \mathbf{b} = \mathbf{{P}}_e^{\downarrow l}\mathbf{{b}}_e^{\downarrow l}.
\end{eqnarray}
The matrix $\mathbf {X}^T_{l}$ and $\mathbf {X}_{l}$ are used to exchange the locations of elements of $\mathbf{{P}}^{\downarrow l}$ and $\mathbf{b}$ respectively, such that $\mathbf{{P}}_{e}^{\downarrow l} = \mathbf{{P}}^{\downarrow l}\mathbf {X}^T_{l}$ and $\mathbf{b}^{\downarrow l}_e = \mathbf {X}_{l} \mathbf{b}$. By multiplying the matrix $\mathbf {X}_{l}$ with $\mathbf{{b}}$, the last $l$ symbols of its each sub-vector $\mathbf{b}_m$ will be moved to the front, i.e.
\begin{eqnarray}\label{eq:12}
\mathbf{b}_{e,m}^{\downarrow l} = [\mathbf b_{m,N-l}\cdots,\mathbf b_{m,N-1},\mathbf b_{m,0},\cdots,\mathbf b_{m,N-l-1}]^T.
\end{eqnarray}
Likewise,
\begin{eqnarray}\label{eq:13}
\mathbf{\bar b}_e^{\downarrow l} = [\mathbf{b}_{e,0}^{\downarrow l};\mathbf{b}_{e,1}^{\downarrow l};\cdots;\mathbf{b}_{e,M-1}^{\downarrow l}] \in \mathbb{C}^{MN\times 1}.
\end{eqnarray}
The effect is similar when multiplying $\mathbf {X}^T_{l}$ with $\mathbf{{P}}^{\downarrow l}$. $\mathbf {X}^T_{l}$ only changes the elements location in $\mathbf{{P}}^{\downarrow l}$.
Substituting (\ref{eq:11}) into (\ref{eq:8}) yields
\begin{eqnarray}\label{eq:14}
\mathbf  {r} =\sum_{l=0}^{L-1} \rho_l \mathbf Z_{l}\mathbf{{P}}_{e}^{\downarrow l}\mathbf{b}_e^{\downarrow l}+\mathbf o_{IBI}+ \mathbf n.
\end{eqnarray}
It can be observed that the non zero elements of $\mathbf{{P}}_{e}^{\downarrow l}$ and  $\mathbf{{P}}$ are very close i.e. the nonzero elements of $\mathbf{{P}}_{e}^{\downarrow l}$ are only delayed by $l$ elements as compared to the elements in $\mathbf{{P}}$. If the non-zero $i^{th}$ row and $k^{th}$ column element of $\mathbf{{P}}$ is $w_{n}$, then the element of $\mathbf{{P}}_{e}^{\downarrow l}$ at the same location will be $w_{n+l}$. Since $N\gg L$, the difference between $w_{n}$ and $w_{n+l}$ is very small as the adjacent elements of the prototype filter are close to each other. In order to show the interference caused by the multipath on the filter distortion, we define $\mathbf{{P}}_{e}^{\downarrow l}$ as follows:
\begin{eqnarray}\label{eq:15}
\mathbf{{P}}_{e}^{\downarrow l} = \mathbf{P} + \Delta\mathbf{P}^{\downarrow l}.
\end{eqnarray}
Eq (\ref{eq:14}) can thus be written as
\begin{eqnarray}\label{eq:16}
\mathbf  {r} = \sum_{l=0}^{L-1} \rho_l \mathbf Z_{l}\mathbf{P}\mathbf{b}_e^{\downarrow l} + \mathbf o_{fd}+ \mathbf o_{IBI}+\mathbf n,
\end{eqnarray}
\begin{figure*}[!t]
	\normalsize
	\setcounter{MYtempeqncnt}{\value{equation}}
	\setcounter{equation}{16}
	\begin{eqnarray}\label{eq:17}
	\!\!\!\mathbf {G}\! =\!\!\!
	\begin{bmatrix}
	\!\!{\mathlarger{\sum_{i=0}^{K-1}\mathbf {W}_i\mathbf {W}_i}} &\!\!\!\!\!\! \mathlarger{\sum_{i=1}^{K-1}\mathbf {W}_{i}\mathbf {W}_{i-1}}&\!\!\!\!\! \cdots&\!\!\!\!\!\!\!\!\! \mathlarger{\sum_{i=K-1}^{K-1}\mathbf {W}_i\mathbf {W}_{i-K+1}}&\!\!\!\!\!\mathbf {0}&\!\!\!\!\!\cdots &\!\!\!\!\!\!\!\!\!\mathbf {0} \\
	\!\!\mathlarger{\sum_{i=1}^{K-1}\mathbf {W}_{i-1}\mathbf {W}_i} &\!\!\!\!\!\! \mathlarger{\sum_{i=0}^{K-1}\mathbf {W}_i\mathbf {W}_i} &\!\!\!\!\! \cdots &\!\!\!\!\!\!\!\!\!\mathlarger{\sum_{i=K-2}^{K-1}\mathbf {W}_{i}\mathbf {W}_{i-K+2}}&\!\!\!\!\! \mathlarger{\sum_{i=K-1}^{K-1}\mathbf {W}_i\mathbf {W}_{i-K+1}}&\!\!\!\!\!\vdots&\!\!\!\!\!\!\!\!\!\mathbf {0} \\
	\!\!\vdots &\!\!\!\!\!\! \vdots &\!\!\!\!\!\ddots &\!\!\!\!\!\!\!\!\! \vdots &\!\!\!\!\!\vdots &\!\!\!\!\! \ddots&\!\!\!\!\!\!\!\!\!\vdots \\
	\!\mathlarger{\sum_{i=K-1}^{K-1}\mathbf {W}_{i-K+1}\mathbf {W}_i} &\!\!\!\!\!\! \mathlarger{\sum_{i=K-2}^{K-1}\mathbf {W}_{i-K+2}\mathbf {W}_i} &\!\!\!\!\! \cdots &\!\!\!\!\!\!\!\!\! \mathlarger{\sum_{i=0}^{K-1}\mathbf {W}_i\mathbf {W}_i} &\!\!\!\!\!\mathlarger{\sum_{i=1}^{K-1}\mathbf {W}_{i}\mathbf {W}_{i-1}} &\!\!\!\!\! \cdots &\!\!\!\!\!\!\!\!\!\mathbf{0}\\
	\!\!\mathbf {0} &\!\!\!\!\!\! \mathlarger{\sum_{i=K-1}^{K-1}\mathbf {W}_{i-K+1}\mathbf {W}_i} &\!\!\!\!\! \cdots  &\!\!\!\!\!\!\!\!\!  \mathlarger{\sum_{i=1}^{K-1}\mathbf {W}_{i-1}\mathbf {W}_i} & \!\!\!\!\! \mathlarger{\sum_{i=0}^{K-1}\mathbf {W}_i\mathbf {W}_i} &\!\!\!\!\!  \cdots &\!\!\!\!\!\!\!\!\! \mathlarger{\sum_{i=K-1}^{K-1}\mathbf {W}_{i-K+1}\mathbf {W}_i}\\
	\!\!\vdots &\!\!\!\!\!\! \vdots &\!\!\!\!\! \ddots  &\!\!\!\!\!\!\!\!\!  \vdots &\!\!\!\!\! \vdots&\!\!\!\!\! \ddots &\!\!\!\!\!\!\!\!\!\vdots\\
	\!\!\mathbf {0} &\!\!\!\!\!\! \mathbf {0} &\!\!\!\!\! \cdots  &\!\!\!\!\!\!\!\!\!  \mathbf {0} &\!\!\!\!\! \mathlarger{\sum_{i=K-1}^{K-1}\mathbf {W}_{i-K+1}\mathbf {W}_i}&\!\!\!\!\!  \cdots&\!\!\!\!\!\!\!\!\! \mathlarger{\sum_{i=0}^{K-1}\mathbf {W}_i\mathbf {W}_i}\\
	\end{bmatrix}\!\!\!
	\end{eqnarray}
	\setcounter{equation}{\value{MYtempeqncnt}}
	\vspace*{4pt}
\end{figure*}
where $\mathbf o_{fd}=\sum_{l=0}^{L-1} \rho_l \mathbf Z_{l}\Delta\mathbf{P}^{\downarrow l}\mathbf{b}_e^{\downarrow l}$ is the interference caused by the filter distortion due to channel multipath effect.
%SECTION II-A-V Receiver Processing
\subsubsection{Receive Processing}
The received signal $\mathbf r$ is first passed through the receive filter bank, represented by the matrix $\mathbf {P}^H$. The output of the receive filter bank becomes
\stepcounter{equation}
\begin{eqnarray}\label{eq:18}
\mathbf  {x} &=& \mathbf{{P}}^{H}\mathbf r, \nonumber \\
&=& \mathbf {G}\sum_{l=0}^{L-1} \rho_l \mathbf Z_{l}\mathbf{b}^{\downarrow l}_e +  \mathbf{P}^{H}(\mathbf o_{fd}+ \mathbf o_{IBI}+ \mathbf n).
\end{eqnarray}
where $\mathbf {G} = \mathbf{P}^{H}\mathbf{P}\in \mathbb{R}^{MN \times MN}$ is the autocorrelation matrix and has a structure as shown in (\ref{eq:17}) with each element being a diagonal sub-matrix of size $N \times N$. We will now analyze the FBMC/QAM system performance with and without an inverse filter at the receiver.
%SECTION II-B CASE 1: No Inverse Filter Matrix
\subsection{Case 1: FBMC/QAM without inverse filter}
The FBMC/QAM system is affected by intrinsic interference introduced by the transmit and receive filters. These interference terms can significantly limit the system performance. In this sub-section, we will derive these interference terms to analyze their impact on the system performance.
%SECTION II-B_I DFT Processing of the filtered signal 
\subsubsection{DFT Processing of the filtered signal}
The signal vector at the output of the receive filter matrix, i.e., $\mathbf{P}^H$ is represented as $\mathbf{x}=[{x_0},{x_1},\cdots,{x_{MN-1}}]^T\in \mathbb{C}^{MN\times 1}$ and is then passed through a serial to parallel converter to split the vector into $M$ segments each of which has $N$ elements to perform $N$-point DFT. The $m^{th}$ segment of the vector $\mathbf {x}$ is represents as $\mathbf{x}_m=[{x}_{mN},{x}_{mN+1},\cdots,\bar{x}_{mN+N-1}]^T\in \mathbb{C}^{N\times1}$ for $m\in 0,1,\cdots,M-1$. The signal is now represented as $\mathbf {x}=[\mathbf{x}_{0},\mathbf{x}_{1},\cdots,\mathbf{x}_{M-1}]\in\mathbb{C}^{N\times M}$ where $\mathbf {x}_{m}=[{x}_{m,0},{x}_{m,1},\cdots,{x}_{m,N-1}]^T\in \mathbb{C}^{N\times 1}$. The signal vector after DFT is represented as follows
\begin{eqnarray}\label{eq:19}
\mathbf {y}_m \!\!\!\!\!\!&=&\!\!\!\!\!\! \mathcal{F}\mathbf {x}_m \in \mathbb{C}^{N \times 1},\nonumber \\
\!\!\!\!\!\!&=&\!\!\!\!\!\! \mathcal{F}\!\!\sum_{i=0}^{M-1}\!\!{\mathbf G}_{m,i}\!\!\sum_{l=0}^{L-1}\!\!\rho_l\mathbf z_{l}\mathbf{b}_{e,i}^{\downarrow l}\!+\! \mathcal{F}\mathbf P_m^H(\mathbf o_{fd}\!+\! \mathbf o_{IBI}\!+\! \mathbf n),
\end{eqnarray}
where ${{\mathbf G}}_{m,i}$ is the $m^{th}$ row and $i^{th}$ column sub-matrices of $\mathbf {G}$. We can show that channel circular convolution property holds in (\ref{eq:19}) and that the channel coefficients and the transmitted signal $\mathbf s_i$ for $i=0,1,...,M$ can be written as point-wise multiplication form in the frequency domain.
We can write $\sum_{l=0}^{L-1}\rho_l\mathbf z_{l} \mathbf{b}_{e,i}^{\downarrow l} = \mathbf H_{cir}\mathbf {b}_i$ in (\ref{eq:19}), where the matrix $\mathbf {H}_{cir}=[\mathbf h_{0}^{\downarrow l},\mathbf h_{1}^{\downarrow l},\cdots,\mathbf h_{L-1}^{\downarrow l}]$ being a $N\times N$ circulant matrix. In general, an $N\times N$ circulant matrix is fully defined by its first $N\times 1$ vector. In our case, $\mathbf H_{cir}$ is determined by $[h_{0}, h_{1},\cdots,h_{L-1},0_{(N-L)\times 1}]^T\in \mathbb{C}^{N \times 1}$ i.e.,
\begin{eqnarray}\label{eq:52}
	\mathbf {H}_{cir}\!\!=\!\!
	\begin{bmatrix}
		\begin{matrix}
			h_0 &\!\! {0} &\!\! \cdots &\!\! {0}&\!\!h_{L-1}&\!\! \cdots&\!\! h_1\\ 
			\vdots &\!\! h_0&\!\! \ddots&\!\! \ddots&\!\! \vdots&\!\! \ddots&\!\! \vdots\\
			h_{L-2}&\!\! \vdots&\!\! \ddots&\!\!\ddots&\!\!\ddots&\!\!0&\!\!h_{L-1}\\
			h_{L-1} &\!\! h_{L-2}&\!\! \cdots&\!\!h_{0}&\!\!0&\!\! \cdots&\vdots\\
			0&\!\!h_{L-1}&\!\!\ddots&\!\! \vdots&\!\! h_{0}&\!\! \cdots&\!\!0\\ 
			\vdots&\!\!\ddots&\!\!\ddots&\!\! \ddots&\!\! \vdots&\!\! \ddots&\!\! \vdots\\
			0&\!\!\cdots&\!\! \cdots&\!\!h_{L-1}&\!\! h_{L-2}&\!\! \cdots&\!\! h_0
		\end{matrix}
	\end{bmatrix},
	%\in\mathbb{C}^{N\times N}
\end{eqnarray}
Also by introducing $\mathcal{F}^H \mathcal{F} = \mathbf I$ in (\ref{eq:19}), we obtain
\begin{eqnarray}\label{eq:53}
	\mathbf  {y}_m \!\!%&=&\!\!\!\!\mathcal{F}\sum_{i=0}^{M-1}{{\mathbf G}}_{m,i}\mathbf H_{cir}\mathbf {b}_i + \mathcal{F}\mathbf P_m^H(\mathbf o_{fd}+ \mathbf o_{IBI}+ \mathbf n)\nonumber \\
	\!\!\!\!&=&\!\!\!\!\mathcal{F}\sum_{i=0}^{M-1}{{\mathbf G}}_{m,i}\mathcal{F}^H \mathcal{F}\mathbf H_{cir}\mathcal{F}^H \mathcal{F}\mathbf {b}_i + \mathcal{F}\mathbf P_m^H(\mathbf o_{fd}\nonumber \\
	&&+ \mathbf o_{IBI}+ \mathbf n).
\end{eqnarray}
Using the circular convolution property (pp.129-130) \cite{sesia2011lte}, we can write $\mathcal{F}\mathbf{H}_{cir}\mathcal {F}^H=\mathbf C$, where $\mathbf C$ is the frequency domain channel coefficients in diagonal matrix form and is given as $\mathbf C=\textrm{diag}[\mathbf C_{0},\mathbf C_{1},\cdots,\mathbf C_{N-1}] \in \mathbb{C}^{N \times N}$. The $n^{th}$ block diagonal element in the frequency response of the channel can be represented as $\mathbf C_{n}=\sum_{l=0}^{L-1} h_{l}e^{-j\frac{2\pi}{N}nl},0\le n\le N$. Also $\mathcal{F}(\mathbf {b}_{i})$ denotes the DFT processing of $\mathbf {b}_{i}$ and according to (\ref{eq:1}), we have $\mathcal{F}(\mathbf {b}_{i}) = \mathbf {s}_i$, by substituting it into (\ref{eq:53}) we get
\begin{eqnarray}\label{eq:20}
\mathbf y_m =\sum_{i=0}^{M-1}{{\mathbf Q}}_{m,i} \mathbf{C} \mathbf {s}_i + \mathcal{F}\mathbf P_m^H(\mathbf o_{fd}+ \mathbf o_{IBI}+ \mathbf n),
\end{eqnarray}
%{\bf Proof:} See Appendix \ref{sec:a0}.\\
%In (\ref{eq:20}), $\mathbf C$ is the frequency domain channel coefficients in diagonal matrix form with its $n^{th}$ diagonal element represented as $\mathbf C_{n}=\mathlarger{\sum_{l=0}^{L-1} h_{l}e^{-j\frac{2\pi}{N}nl}},0\le n\le N$ and 
where $\mathbf Q_{m,i}= \mathcal{F}{{\mathbf G}}_{m,i}\mathcal{F}^H$ has the following property
\begin{eqnarray}\label{eq:21}
\mathbf Q_{m,i} = \left\{
    \begin{array}{lcl}
    \mathbf I + \Delta \mathbf Q_{m,m} \quad \textrm{for} \quad  i=m\\
    \Delta \mathbf Q_{m,i}  \quad \quad \quad \textrm{for} \quad i\neq m
    \end{array}
\right.,
\end{eqnarray}
Note that $\Delta \mathbf Q \in \mathbb C^{MN \times MN}$ % = \mathbf I - \mathbf Q\in \mathbb{C}^{MN \times MN}$
denotes the interference coefficient matrix that determines the magnitude of intrinsic interference in the received signal block. Using (\ref{eq:21}), we can write (\ref{eq:20}) as follows
\begin{eqnarray}\label{eq:22}
\mathbf y_m \!\!\!\!&=&\!\!\!\! (\mathbf I + \Delta \mathbf Q_{mm})\mathbf C \mathbf s_m\!\!\!+\!\!\!\sum_{i=0,i\neq m}^{M-1}\!\!\!\Delta \mathbf Q_{m,i}\mathbf{C} \mathbf {s}_i \nonumber \\
&&+ \mathcal{F}\mathbf P_m^H(\mathbf o_{fd}+ \mathbf o_{IBI}+ \mathbf n),\nonumber \\
\!\!\!\!&=&\!\!\!\!\mathbf C\mathbf s_m \!+\! \Delta \mathbf Q_{mm}\mathbf C \mathbf s_m\!\!+\!\!\!\!\sum_{i=0,i\neq m}^{M-1}\!\!\!\!\!\Delta \mathbf Q_{m,i}\mathbf{C} \mathbf {s}_i \nonumber \\
&&+ \mathcal{F}\mathbf P_m^H(\mathbf o_{fd}+ \mathbf o_{IBI}+ \mathbf n).
\end{eqnarray}
%SECTION II-B-II Channel Equalization
\subsubsection{Channel Equalization}
We represent one tap channel equalizer as a diagonal matrix $\mathbf E$ and is applied to the signal $\mathbf {y}_m$ as follows
\begin{eqnarray}\label{eq:23}
\mathbf {\hat s}_m \!\!\!\!&=&\!\!\!\! \mathbf E \mathbf {y}_m, \nonumber \\
\!\!\!\!&=&\!\!\!\!\mathbf E\mathbf C\mathbf s_m \!\!+\!\! \mathbf E\Delta \mathbf Q_{mm}\mathbf C \mathbf s_m\!\!+\!\!\mathbf E\!\!\!\!\!\sum_{i=0,i\neq m}^{M-1}\!\!\!\!\!\Delta \mathbf Q_{m,i}\mathbf{C} \mathbf {s}_i \nonumber \\
&&+ \mathbf E\mathcal{F}\mathbf P_m^H(\mathbf o_{fd}+ \mathbf o_{IBI}+ \mathbf n),
\end{eqnarray}
Let us now assume $\mathbf E$ to be either ZF or MMSE i.e. the two most popular linear channel equalizers.
\begin{eqnarray}\label{eq:24}
\mathbf {E} = {\mathbf C}^H(\mathbf C \mathbf C^H + \nu\sigma^2/\delta^2\mathbf{I})^{-1},
\end{eqnarray}
where $\nu = 0$ for ZF while $\nu = 1$ is for MMSE. We can now write (\ref{eq:23}) as
\begin{eqnarray}\label{eq:25}
\mathbf {\hat s}_m\!\!\!\!&=&\!\!\!\!\boldsymbol{\beta} \mathbf s_m \!+\! \mathbf E\Delta \mathbf Q_{mm}\mathbf C \mathbf s_m+\!\!\!\!\!\sum_{i=0,i\neq m}^{M-1}\!\!\!\!\!\mathbf E\Delta \mathbf Q_{m,i}\mathbf{C} \mathbf {s}_i \nonumber \\
&&+ \mathbf E\mathcal{F}\mathbf P_m^H(\mathbf o_{fd}+ \mathbf o_{IBI}+ \mathbf n),
\end{eqnarray}
where $\boldsymbol{\beta} = \mathbf E\mathbf C$ is a diagonal matrix with its $n^{th}$ diagonal element being defined as
\begin{eqnarray}\label{eq:26}
{\beta}_{n} = {E}_{n} {C}_{n}= \frac{|E_{n}|^2}{|E_{n}|^2 + \nu\sigma^2/\delta^2},
\end{eqnarray}
The estimated signal $\mathbf{\hat s}_m$ can now be expressed as follows
\begin{eqnarray}\label{eq:27}
\mathbf {\hat s}_m=\underbrace{\mathbf s_m}_{\textrm{Desired Signal}}+\underbrace{(\mathbf{I}-\boldsymbol{\beta})\mathbf s_m}_{\textrm{MMSE Estimation Bias}}\nonumber \\+ \underbrace{\mathbf E\Delta \mathbf Q_{mm}\mathbf C \mathbf s_m}_{\textrm{ICI}}+\underbrace{\sum_{i=0,i\neq m}^{M-1}\mathbf E\Delta \mathbf Q_{m,i}\mathbf{C} \mathbf {s}_i}_{\textrm{ISI}} \nonumber \\+ \!\!\!\!\!\!\!\!\!\!\underbrace{\mathbf E\mathcal{F}\mathbf P_m^H \mathbf o_{fd}}_{\textrm{Filter Distortion by Multipath}}\!\!\!\!\!+ \underbrace{\mathbf E\mathcal{F}\mathbf P_m^H\mathbf o_{IBI}}_{\textrm{IBI by Multipath}}+ \underbrace{\mathbf E\mathcal{F}\mathbf P_m^H\mathbf n}_{\textrm{Noise}}.
\end{eqnarray}
Note that the estimation bias error $(\mathbf I-\boldsymbol\beta) $ is an effect of compromising the interference and noise of the MMSE equalizer. However, $(\mathbf I -\boldsymbol\beta) = \mathbf 0$ when the ZF receiver is used. 
%SECTION II-C CASE 2: Inverse Filter Matrix
\subsection{Case 2: FBMC/QAM with inverse filter} \label{sp_mtx}
\begin{figure}[th]
\centering
\includegraphics[scale=0.5]{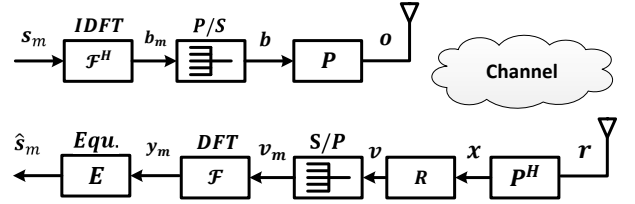}
\caption{Block diagram of FBMC/QAM system with Inverse Filter}
\label{fig2}
\end{figure}

\begin{figure}[th]
	\centering
	\includegraphics[scale=0.3]{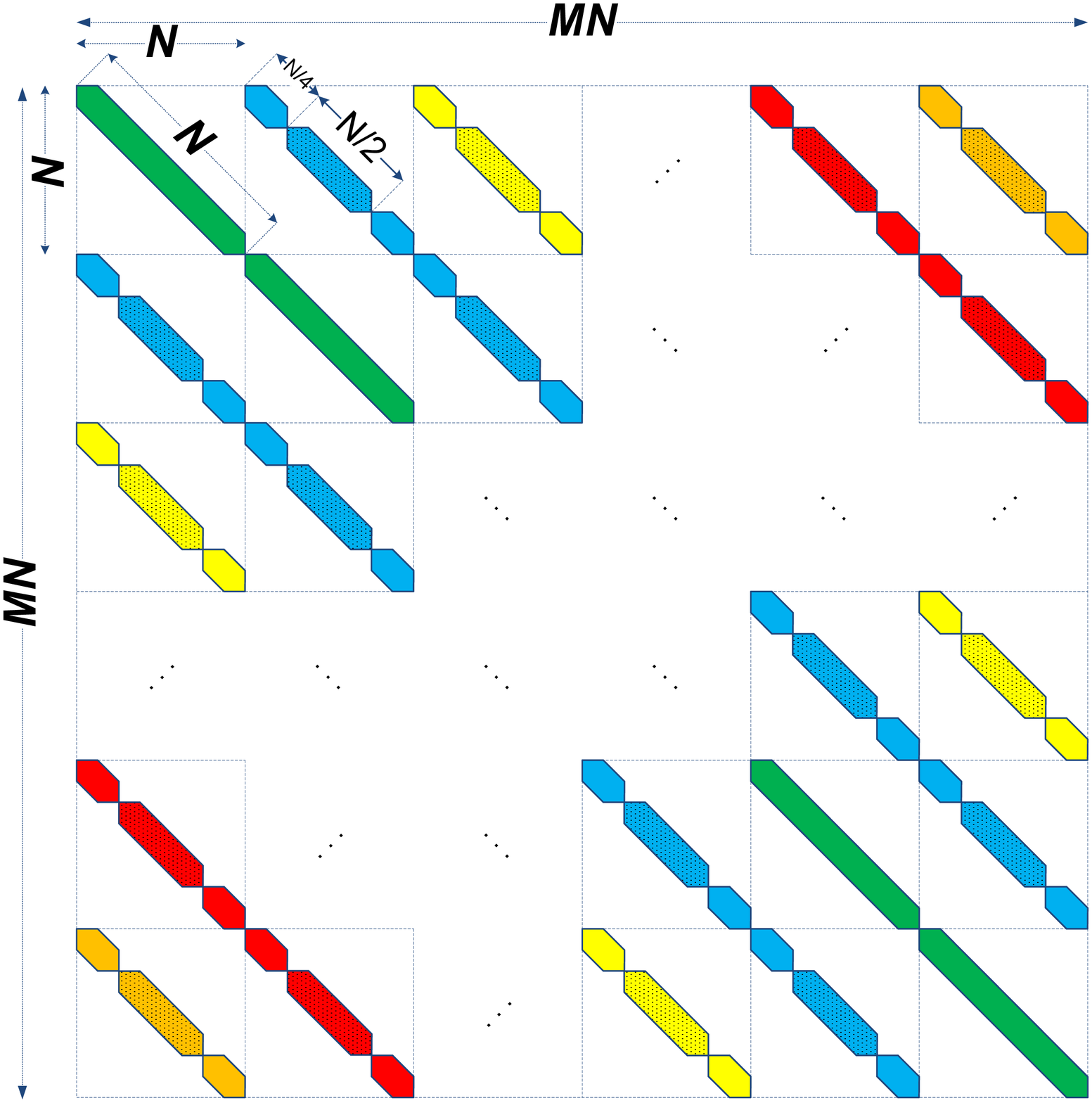}
	\caption{Structure of inverse filter matrix $R$}
	\label{fig3}
\end{figure}

We can see from (\ref{eq:27}) that the transmitted signal $\mathbf s_m$ is accompanied by ICI and ISI (intrinsic interference) terms along with interferences caused by the multipath channel and the noise. We can overcome the intrinsic interference by introducing an inverse filter matrix $\mathbf R$ at the receiver as shown in Fig. \ref{fig2}.
Let the inverse filter matrix be defined as the inverse of the autocorrelation matrix $\mathbf G$ defined in (\ref{eq:17}) i.e. $\mathbf R= \mathbf G^{-1}\in \mathbb{R}^{MN\times MN}$. %It is worth mentioning that since the coefficients in the autocorrelation matrix $\mathbf G$ are constant, we can calculate the inverse filter matrix off-line. Secondly, 
Since the autocorrelation matrix $\mathbf G$ is a band diagonal matrix, the inverse of the band diagonal matrix will result in a sparse matrix that consists of diagonal sub-matrices as shown in Fig. \ref{fig3}. The sparse structure of the inverse filter can lead to a low complex deconvolution process at the receiver. It should be noted that each off diagonal sub-matrix in $\mathbf R$ has negligible middle $N/2$ diagonal elements represented as dotted sections in Fig. \ref{fig3}. An arbitrary number of elements in the range of $[0, N/2]$ can be replaced here by zero to reduce the complexity. Let the elements of the dotted section considered in the complexity analysis be defined as $\eta$ %However, the middle $N/4$ elements represented as dotted section will be considered when comparing the complexity of the system in Sec. \ref{comp_ana}. We will later investigate the impact of neglecting these elements on the BER performance of the system in Sec. \ref{ber}.
i.e., if %s introduce a variable $\eta$ representing the elements considered in the dotted sections to investigate its impact on the complexity analysis i.e., 
$\eta\!=\!0$, all $N/2$ diagonal elements are considered, $\eta=0.5$ represent $N/4$ diagonal elements in the range of $[3N/8,5N/8]$  are considered, whereas $\eta\!=\!1$ means none of the middle $N/2$ diagonal elements are considered %in each off-diagonal sub-matrix in $\mathbf R$ 
in the complexity analysis given in Sec. \ref{comp_ana}.
\\Using (\ref{eq:18}), the output of the inverse filter matrix $\mathbf R$ at the receiver side can be written as follows
\begin{eqnarray}\label{eq:28}
\mathbf v_m \!\!=\!\!\sum_{i=0}^{M-1}{{\mathbf U}}_{m,i} \sum_{l=0}^{L-1} \rho_l \mathbf z_{l}\mathbf{b}_{e,i}^{\downarrow l} + \mathbf R_m\mathbf P_m^H(\mathbf o_{fd}+ \mathbf o_{IBI}+ \mathbf n),
\end{eqnarray}
where ${{\mathbf R}}_{m}\in \mathbb{C}^{N \times MN}$ is the $m^{th}$ row of sub-matrices of matrix $\mathbf {R}$, while $\mathbf U_{m,i} = \sum_{j=0}^{M-1}\mathbf R_{m,j}\mathbf G_{j,i}\in \mathbb{R}^{N \times N}$ and has the following property
\begin{eqnarray}\label{eq:29}
\mathbf U_{m,i} = \left\{
\begin{array}{lcl}
\mathbf I_{N \times N} \quad \textrm{for} \quad  i=m\\
\mathbf 0 \quad \quad \quad \textrm{for} \quad i\neq m
\end{array}
\right.,
\end{eqnarray}
Eq. (\ref{eq:28}) can now be written as follows
\begin{eqnarray}\label{eq:30}
\mathbf v_m \!\!=\!\!\sum_{l=0}^{L-1} \rho_l \mathbf z_{l}\mathbf{b}_{e,m}^{\downarrow l} + \mathbf R_m\mathbf P_m^H(\mathbf o_{fd}+ \mathbf o_{IBI}+ \mathbf n).
\end{eqnarray}
%SECTION II-C-I DFT Processing of the filtered signal 
\subsubsection{DFT Processing of the filtered signal}
The signal after DFT processing is now represented as follows
\begin{eqnarray}\label{eq:31}
\mathbf  {y}_m = \mathbf C\mathbf s_m + \mathcal{F}\mathbf R_m\mathbf P_m^H(\mathbf o_{fd}+ \mathbf o_{IBI}+ \mathbf n).
\end{eqnarray}
where $\mathbf C\mathbf {s}_m=\mathcal{F}\sum_{l=0}^{L-1}\rho_l\mathbf z_{l} \mathbf{b}_{e,m}^{\downarrow l}$. %as derived in Appendix \ref{sec:a0}.
%SECTION II-C-II Channel Equalization
\subsubsection{Channel Equalization}
The estimated symbol $\mathbf {\hat s}_m$ after equalization can be expressed as follows
\begin{eqnarray}\label{eq:32}
\mathbf {\hat s}_m \!\!\!\!\!&=&\!\!\!\!\! \mathbf E \mathbf {y}_m,\nonumber \\
\!\!\!\!\!&=&\!\!\!\!\!\boldsymbol{\beta} \mathbf s_m + \mathbf E\mathcal{F}\mathbf R_m\mathbf P_m^H(\mathbf o_{fd}+ \mathbf o_{IBI}+ \mathbf n),
\end{eqnarray}
where $\boldsymbol{\beta} \!\!=\!\! \mathbf E\mathbf C$ is a diagonal matrix with its $n^{th}$ diagonal element represented as (\ref{eq:26}).
The estimated signal $\mathbf{\hat s}_m$ can now be expressed as follows:
\begin{eqnarray}\label{eq:33}
\mathbf {\hat s}_m=\underbrace{\mathbf s_m}_{\textrm{Desired Signal}}+\underbrace{(\mathbf{I}-\boldsymbol{\beta})\mathbf s_m}_{\textrm{MMSE Estimation Bias}}\nonumber \\+\!\!\!\!\!\! \underbrace{\mathbf E\mathcal{F}\mathbf R_m\mathbf P_m^H \mathbf o_{fd}}_{\textrm{Filter Distortion by Multipath}}\!\!\! + \underbrace{\mathbf E\mathcal{F}\mathbf R_m\mathbf P_m^H\mathbf o_{IBI}}_{\textrm{IBI by Multipath}}+ \underbrace{\mathbf E\mathcal{F}\mathbf R_m\mathbf P_m^H\mathbf n}_{\textrm{Noise}}.
\end{eqnarray}
\indent  
As we can see from (\ref{eq:33}) that the transmitted signal $\mathbf s_m$ is free from ICI and ISI terms as compared to the case with no inverse filter. However, the use of inverse filter matrix $\mathbf R$ enhances the interferences caused by the multipath channel and noise as shown in (\ref{eq:33}). Therefore, in what follows, we will investigate the interference and noise power to analyze the usefulness of inverse filter matrix at the receiver.
%SECTION III Interference Analysis
\section{Interference Analysis}
Although the use of inverse filter matrix removes ISI and ICI, however, the interference caused by the multipath channel and noise is enhanced due to the use of inverse filter matrix $\mathbf R$. Therefore we need to investigate the interference and noise power enhancement due to the inverse filter at the receiver. This provides deep insights and useful guidelines for receiver design in the FBMC/QAM system.
%SECTION III-A Interference / noise power in case of no inverse filter
\subsection {Interference / noise power in case of no inverse filter}
As we can see from (\ref{eq:27}) that in case of no inverse filter, the estimated symbol is accompanied with MMSE estimation bias, interference terms like ICI, ISI, filter distortion and IBI due to multipath channel and noise i.e.,
\begin{eqnarray}\label{eq:34}
\mathbf {\hat s}_m=\underbrace{\mathbf s_m}_{\textrm{Desired Signal}}\!\!\!+\underbrace{\boldsymbol{\psi}_{resd,m}}_{\textrm{MMSE Estimation Bias}}\!\!\!+ \underbrace{\boldsymbol{\psi}_{ICI,m}}_{\textrm{ICI}}+\underbrace{\boldsymbol{\psi}_{ISI,m}}_{\textrm{ISI}} \nonumber \\+\!\!\!\!\!\! \underbrace{\boldsymbol{\psi}_{fd,m}}_{\textrm{Filter Distortion by Multipath}}\!\!\!+ \underbrace{\boldsymbol{\psi}_{IBI,m}}_{\textrm{IBI by Multipath}}+ \underbrace{\boldsymbol{\psi}_{noise,m}}_{\textrm{Noise}}.
\end{eqnarray}
\indent The MSE of the $n$-th modulation symbol estimation in the $m$-th FBMC/QAM symbol can be derived as
\begin{eqnarray}\label{eq:35}
\gamma_{tot,m,n}\!\!\!\! &=&\!\!\!\! \mathcal {E}||{\hat s}_{m,n}- {s}_{m,n}||^2 \nonumber \\
\!\!\!\!&=& \!\!\!\!\mathcal {E}\big[ \|\boldsymbol{\psi}_{resd,m}\|_n^2+\|\boldsymbol{\psi}_{ICI,m}\|_n^2+ \|\boldsymbol{\psi}_{ISI,m}\|_n^2\!\nonumber\\\!\!&&+\! \|\boldsymbol{\psi}_{fd,m}\|_n^2\!\!+\!\! \|\boldsymbol{\psi}_{IBI,m}\|_n^2\!\!+\!\!\|\boldsymbol{\psi}_{noise,m}\|_n^2\big].
\end{eqnarray}
%SECTION III-A-I MSE of signal estimation bias
\subsubsection{MSE of signal estimation bias}
The desired signal estimation bias is caused by the MMSE receiver since it minimizes the MSE between the transmitted and received signal. This leads to residual interference in the estimated signal. From (\ref{eq:35}) and (\ref{eq:27}), we can write the variance of the signal estimation bias as
\begin{eqnarray}\label{eq:54}
	\boldsymbol\gamma_{resd,m,n}\!\!\!&=&\!\!\!\mathcal {E}\|\boldsymbol{\psi}_{resd,m}\|_n^2 = \mathcal {E}\{\|(\mathbf I -\boldsymbol\beta)\mathbf s_m\|_n^2\}\nonumber \\
	\!\!&=&\!\!\delta^2(I -\beta_n)^2,
\end{eqnarray}
As $\mathcal{E}\{\|s_{m,n}\|^2\}\!\!=\!\! \delta^2$ and according to (\ref{eq:26}), ${\beta_n} = \frac{|C_n|^2}{|C_n|^2 + \nu\sigma^2/\delta^2}$. Substituting ${\beta_n}$ in (\ref{eq:54}) yields
\begin{eqnarray}\label{eq:55}
	\gamma_{resd,m,n} \!\!\!\!&=&\!\!\!\! \delta^2(I -\beta_n)^2=\delta^2(I -2\beta_n+\beta_n^2], \nonumber \\
	\!\!\!\!&=&\!\!\!\!\delta^2\Big[I -\frac{2|C_n|^2}{|C_n|^2 + \nu\sigma^2/\delta^2}+\frac{|C_n|^4}{(|C_n|^2 + \nu\sigma^2/\delta^2)^2}\Big],\nonumber \\
	%&=&\delta^2\Big[\frac{(|C_n|^2 + \nu\sigma^2/\delta^2)^2-2|C_n|^2\{|C_n|^2 + \nu\sigma^2/\delta^2\}+|C_n|^4}{(|C_n|^2 + \nu\sigma^2/\delta^2)^2}\Big]\nonumber \\
	%&=&\delta^2\Big[\frac{(|C_n|^2 + \nu\sigma^2/\delta^2)^2-|C_n|^4-2|C_n|^2 \nu\sigma^2/\delta^2}{(|C_n|^2 + \nu\sigma^2/\delta^2)^2}\Big]\nonumber \\
	%&=&\delta^2\bigg[\frac{|C_n|^4 + \nu^2\sigma^4/\delta^4+2|C_n|^2 \nu\sigma^2/\delta^2-|C_n|^4-2|C_n|^2 \nu\sigma^2/\delta^2}{(|C_n|^2 + \nu\sigma^2/\delta^2)^2}\bigg]\nonumber \\
	%&=&\delta^2\Big[\frac{\nu^2\sigma^4/\delta^4}{(|C_n|^2 + \nu\sigma^2/\delta^2)^2}\Big]\nonumber \\
	\!\!\!\!&=&\!\!\!\!\delta^2\bigg[\frac{\nu^2\sigma^4}{(\delta^2|C_n|^2 + \nu\sigma^2)^2}\bigg].
\end{eqnarray}
%\begin{eqnarray}\label{eq:36}
%\boldsymbol\gamma_{resd,m,n}&=&\mathcal {E}\|\boldsymbol{\psi}_{resd,m}\|_n^2 = \mathcal {E}\|(\mathbf I -\boldsymbol\beta)\mathbf s_m\|_n^2,\nonumber \\
%&=&\delta^2\bigg[\frac{\nu^2\sigma^4}{(\delta^2|C_n|^2 + \nu\sigma^2)^2}\bigg].
%\end{eqnarray}
%{\bf Proof:} See Appendix \ref{sec:a1}.\\
Apparently, when the ZF receiver is adopted, $\gamma_{resd,m,n} = 0 $ since $\nu = 0$. However, the ZF equalization leads to noise enhancement unlike MMSE receivers.
%%SECTION III-A-II MSE of ICI
\subsubsection{MSE of ICI}
We can write the variance of the ICI from (\ref{eq:35}) and (\ref{eq:27}) as
\begin{eqnarray}\label{eq:56}
	\boldsymbol\gamma_{ICI,m}\!\!\!\!&=&\!\!\!\!\mathcal {E}\|\boldsymbol{\psi}_{ICI,m}\|^2 = \mathcal {E}\|\mathbf E\Delta \mathbf Q_{mm}\mathbf C \mathbf s_m\|^2,\nonumber \\
	\!\!\!\!&=&\!\!\!\! \mathcal {E}[\mathbf E\Delta\mathbf Q_{mm}\mathbf C \mathbf s_m\mathbf s^H_m\mathbf C^H\Delta\mathbf Q^H_{m,m}\mathbf E^H],
\end{eqnarray}
As $\mathcal{E}\{\|s_{m,n}\|^2\}\!\!\!=\!\!\! \delta^2$ and from (\ref{eq:21}), we know that $\Delta\mathbf Q_{m,m}\!\!=\!\!\mathbf Q_{m,m}\!\!-\!\mathbf I$ for $i\!\!=\!\!m$, we can thus reformulate the above equation as
\begin{eqnarray}\label{eq:57}
	\boldsymbol\gamma_{ICI,m}\!\!\!\!&=&\!\!\!\! \delta^2\mathbf E\Delta\{\mathbf Q_{m,m}-\mathbf I\}\mathbf C\mathbf C^H\Delta\{\mathbf Q_{m,m}-\mathbf I\}^H\mathbf E^H,\nonumber \\
	\!\!\!\!&=&\!\!\!\!\delta^2\mathbf E\mathbf Q_{m,m}\mathbf C\mathbf C^H\mathbf Q^H_{m,m}\mathbf E^H-\mathbf E\mathbf Q_{m,m}\mathbf C\mathbf C^H\mathbf E^H\nonumber \\
	&&-\mathbf E\mathbf C\mathbf C^H\mathbf Q^H_{m,m}\mathbf E^H+\mathbf E\mathbf C\mathbf C^H\mathbf E^H,%\nonumber \\
	%&=& \delta^2[|\mathbf E|^2|\mathbf C|^2|\mathbf Q_{m,m}|^2-|\mathbf E|^2|\mathbf C|^2\mathbf Q_{m,m}-|\mathbf E|^2|\mathbf C|^2\mathbf Q^H_{m,m}+|\mathbf E|^2|\mathbf C|^2]\nonumber \\
	%&=& \delta^2|\mathbf E|^2|\mathbf C|^2[|\mathbf Q_{m,m}|^2-\mathbf Q_{m,m}-\mathbf Q^H_{m,m}+\mathbf I]\nonumber \\
	%&=& \delta^2|\mathbf E|^2|\mathbf C|^2 \boldsymbol \alpha_{ICI}
\end{eqnarray}
Taking the $n^{th}$ diagonal element of (\ref{eq:57}), we have
\begin{eqnarray}\label{eq:58}
	\boldsymbol\gamma_{ICI,m,n}\!\!\!\!&=&\!\!\!\!\delta^2\|\mathbf E\mathbf Q_{m,m}\mathbf C\mathbf C^H\mathbf Q^H_{m,m}\mathbf E^H\|_n\!\!-\!\!\|\mathbf E\mathbf Q_{m,m}\mathbf C\mathbf C^H\nonumber\\
	&&\mathbf E^H\|_n-\|\mathbf E\mathbf C\mathbf C^H\mathbf Q^H_{m,m}\mathbf E^H\|_n\!\!+\!\!\|\mathbf E\mathbf C\mathbf C^H\mathbf E^H\|_n, \nonumber \\
	\!\!\!&=&\!\!\! \delta^2|E_n|^2|C_n|^2 \|\mathbf Q_{m,m}\mathbf Q^H_{m,m}\|_n-|E_n|^2|C_n|^2 \nonumber \\
	&&\|\mathbf Q_{m,m}\|_n
	\!\!-\!\!|E_n|^2|C_n|^2 \|\mathbf Q^H_{m,m}\|_n \!\!+\!\! |E_n|^2|C_n|^2,\nonumber \\
	\!\!\!&=&\!\!\! |E_n|^2|C_n|^2\alpha_{ICI,n}.
\end{eqnarray}
where $\alpha_{ICI,n}=\|\delta^2\mathbf Q_{m,m}\mathbf Q^H_{m,m}- \mathbf Q_{m,m}-\mathbf Q^H_{m,m} +I_{N\times N}\|_n$.
%\begin{eqnarray}\label{eq:37}
%\boldsymbol\gamma_{ICI,m,n}&=&\mathcal {E}\|\boldsymbol{\psi}_{ICI,m}\|_n^2 = \mathcal {E}\|\mathbf E\Delta \mathbf Q_{mm}\mathbf C \mathbf s_m\|_n^2,\nonumber \\
%&=&|E_n|^2|C_n|^2\alpha_{ICI,n}.
%\end{eqnarray}
%where $\alpha_{ICI,n}=\|\delta^2\mathbf Q_{m,m}\mathbf Q^H_{m,m}- \mathbf Q_{m,m}-\mathbf Q^H_{m,m} +\mathbf I_{N\times N}\|_n$ \\
%{\bf Proof:} See Appendix \ref{sec:a3}.
%%SECTION III-A-III MSE of ISI
\subsubsection{MSE of ISI}
%The variance of inter-symbol interference is
We can write the variance of the ISI from (\ref{eq:35}) and (\ref{eq:27}) as
\begin{eqnarray}\label{eq:59}
	\boldsymbol\gamma_{ISI,m}\!\!\!&=&\!\!\!\mathcal {E}\|\boldsymbol{\psi}_{ISI,m}\|^2 = \mathcal {E}\|\sum_{i=0,i\neq m}^{M-1}\mathbf E\Delta\mathbf Q_{m,i}\mathbf{C} \mathbf {s}_i\|^2,\nonumber \\
	&=&\!\!\! \mathcal {E}[\sum_{i=0,i\neq m}^{M-1}\!\!\!\mathbf E\Delta \mathbf Q_{m,i}\mathbf{C} \mathbf {s}_i\mathbf {s}^H_i\mathbf{C}^H\Delta \mathbf Q^H_{m,i}\mathbf E^H],
\end{eqnarray}
As $\mathcal{E}\{\|s_{m,n}\|^2\}\!\!=\!\! \delta^2$ and from (\ref{eq:21}), we know that $\Delta\mathbf Q_{m,i}=\mathbf Q_{m,i}$ for $i\ne m$, we can thus write the above equation as
\begin{eqnarray}\label{eq:60}
	\boldsymbol\gamma_{ISI,m}\!\!\!\!&=&\!\!\!\! \delta^2\!\!\!\sum_{i=0,i\neq m}^{M-1}\mathbf E\mathbf Q_{m,i}\mathbf{C}\mathbf{C}^H \mathbf Q^H_{m,i}\mathbf E^H,
	%&=& \delta^2|\mathbf E|^2|\mathbf C|^2 \sum_{i=0,i\neq m}^{M-1}|\mathbf Q_{m,i}|^2\nonumber \\
	%&=& \delta^2|\mathbf E|^2|\mathbf C|^2 \boldsymbol \alpha_{ISI}
\end{eqnarray}
Taking the $n^{th}$ diagonal element of (\ref{eq:60}), we obtain
\begin{eqnarray}\label{eq:61}
	\gamma_{ISI,m,n}\!\!\!\!&=&\!\!\!\! \delta^2\|\sum_{i=0,i\neq m}^{M-1}\mathbf E\mathbf Q_{m,i}\mathbf{C}\mathbf{C}^H \mathbf Q^H_{m,i}\mathbf E^H\|_n,\nonumber\\
	\!\!\!\!&=&\!\!\!\! \delta^2|E_n|^2|C_n|^2\alpha_{ISI,n}.
\end{eqnarray}
where $\alpha_{ISI,n}=\|\!\!\!\mathlarger{\sum_{i=0,i\neq m}^{M-1}}\!\!\!\mathbf Q_{m,i} \mathbf Q^H_{m,i}\|_n$.
%\begin{eqnarray}\label{eq:38}
%\boldsymbol\gamma_{ISI,m,n}&=&\mathcal {E}\|\boldsymbol{\psi}_{ISI,m}\|_n^2 = \mathcal {E}\|\sum_{i=0,i\neq m}^{M-1}\mathbf E\Delta\mathbf Q_{m,i}\mathbf{C} \mathbf {s}_i\|_n^2,\nonumber \\
%&=& \delta^2|E_n|^2|C_n|^2\alpha_{ISI,n}.
%\end{eqnarray}
%where $\alpha_{ISI,n}=\|\mathlarger{\sum_{i=0,i\neq m}^{M-1}}\mathbf Q_{m,i} \mathbf Q^H_{m,i}\|_n$\\
%{\bf Proof:} See Appendix \ref{sec:a4}.
%SECTION III-A-IV MSE of distorted filter
\subsubsection{MSE of Filter Distortion due to multipath channel}
%The MSE caused by the filter distortion due to channel multipath effect is
We can write the variance of the interference caused by filter distortion due to multipath channel from (\ref{eq:35}) and (\ref{eq:27}) as
\begin{eqnarray}\label{eq:62}
	\boldsymbol\gamma_{fd,m}\!\!\!&=&\!\!\!\mathcal {E}\|\boldsymbol{\psi}_{fd,m}\|^2 = \mathcal {E}\|\mathbf E\mathcal{F}\mathbf P_m^H \mathbf o_{fd}\|^2,\nonumber \\
	\!\!\!&=&\!\!\! \mathcal {E}[\mathbf E\mathcal{F}\mathbf P_m^H \mathbf o_{fd}\mathbf o^H_{fd}\mathbf P_m\mathcal{F}^H\mathbf E^H],\nonumber \\
	\!\!\!&=&\!\!\! \mathbf E\mathcal{F}\mathbf P_m^H \mathcal{E}[\mathbf o_{fd}\mathbf o^H_{fd}]\mathbf P_m\mathcal{F}^H\mathbf E^H,\nonumber \\
	\!\!\!&=&\!\!\! \mathbf E\mathcal{F}\mathbf P_m^H \boldsymbol\alpha_{fd}\mathbf P_m\mathcal{F}^H\mathbf E^H,
\end{eqnarray}
Using (\ref{eq:16}), we can determine $\boldsymbol\alpha_{fd}=\mathcal{E}[\mathbf o_{fd}\mathbf o^H_{fd}]$ as follows
\begin{eqnarray}\label{eq:63}
	\boldsymbol\alpha_{fd}\!\!\!\!&=&\!\!\!\! \mathcal {E}\Big[\Big\{ \sum_{l=0}^{L-1} \rho_l \mathbf Z_{l}\Delta\mathbf{P}^{\downarrow l}\mathbf{b}_e^{\downarrow l}\Big\}\{ \sum_{l=0}^{L-1} \rho_l \mathbf Z_{l}\Delta\mathbf{P}^{\downarrow l}\mathbf{b}_e^{\downarrow l}\Big\}^H\Big],\nonumber \\
	\!\!\!\!&=&\!\!\!\! \sum_{l=0}^{L-1}\rho^2_l\mathcal {E}[\mathbf Z_{l}\Delta\mathbf{P}^{\downarrow l}\mathbf{b}_e^{\downarrow l}\mathbf{b}_e^{\downarrow lH}\Delta\mathbf{P}^{\downarrow lH}\mathbf Z^H_{l}],%\nonumber \\
	%&=& \delta^2\sum_{l=0}^{L-1}\rho^2_l\mathcal {E}[\mathbf Z_{l}\Delta\mathbf{P}^{\downarrow l}\Delta\mathbf{P}^{\downarrow lH}\mathbf Z^H_{l}]
\end{eqnarray}
From (\ref{eq:4}) and (\ref{eq:6}), $\mathcal{E}\{\mathbf Z_l\mathbf Z_l^H\}\!\!=\!\!1$ since $z_l \in \mathbb{C}\mathcal{N}(0,1)$ also we know that $\mathcal{E}\{\mathbf{b}_e^{\downarrow l}\mathbf{b}_e^{\downarrow lH}\}=\delta^2$, consequently
\begin{eqnarray}\label{eq:64}
	\boldsymbol\alpha_{fd}\!\!\!\!&=&\!\!\!\!\delta^2 \sum_{l=0}^{L-1}\rho^2_l\textrm{Tr}\{\Delta\mathbf{P}^{\downarrow l}\Delta\mathbf{P}^{\downarrow lH}\},\nonumber \\
	\!\!\!\!&=&\!\!\!\!\delta^2\sum_{l=0}^{L-1}\rho^2_l T^{\downarrow l},
\end{eqnarray}
where $T^{\downarrow l}\!\!=\!\! \textrm{Tr}[\Delta\mathbf{P}^{\downarrow l}\Delta\mathbf{P}^{\downarrow lH}]$. %Substituting Eq (\ref{eq:72}) in Eq (\ref{eq:39}), we get
Since $T^{\downarrow l}$ is a scalar value, therefore $\boldsymbol\alpha_{fd}$ is also a scalar value. Now substituting (\ref{eq:64}) into (\ref{eq:62}), yields
\begin{eqnarray}\label{eq:65}
	\boldsymbol \gamma_{fd,m}= \alpha_{fd}\mathbf E\mathcal{F}\mathbf P_m^H\mathbf P_m\mathcal{F}^H\mathbf E^H,
\end{eqnarray}
By taking the $n^{th}$ diagonal element of $\boldsymbol \gamma_{fd,m}$, we have
\begin{eqnarray}\label{eq:66}
	\gamma_{fd,m,n}\!\!\!&=&\!\!\!\alpha_{fd}\|\mathbf E\mathcal{F}\mathbf P_m^H\mathbf P_m\mathcal{F}^H\mathbf E^H\|_n,\nonumber\\
	\!\!\!&=&\!\!\!\alpha_{fd}|E_n|^2.
\end{eqnarray}
where $\|\mathcal{F}\mathbf P_m^H\mathbf P_m\mathcal{F}^H\|_n=\|\mathbf I_{N\times N}\|_n$.
%\begin{eqnarray}\label{eq:39}
%\boldsymbol\gamma_{fd,m,n}&=&\mathcal {E}\|\boldsymbol{\psi}_{fd,m}\|_n^2 = \mathcal {E}\|\mathbf E\mathcal{F}\mathbf P_m^H \mathbf o_{fd}\|_n^2,\nonumber \\
%&=&\alpha_{fd}|E_n|^2.
%\end{eqnarray}
%where $\alpha_{fd}=\delta^2 \mathlarger{\sum_{l=0}^{L-1}}\rho^2_l\textrm{Tr}\{\Delta\mathbf{P}^{\downarrow l}(\Delta\mathbf{P}^{\downarrow l})^H\}$.\\
%{\bf Proof:} See Appendix \ref{sec:a5}.
%%SECTION III-A-V MSE of IBI
\subsubsection{MSE of IBI}
Let us consider the case when we have inter-block interference due to the lack of guard time. %The variance of interference caused by IBI in this case is
We can write the variance of the interference caused by IBI from (\ref{eq:35}) and (\ref{eq:27}) as
\begin{eqnarray}\label{eq:67}
	\boldsymbol\gamma_{IBI,m}\!\!\!\!&=&\!\!\!\!\mathcal {E}\|\boldsymbol{\psi}_{IBI,m}\|^2 = \mathcal {E}\|\mathbf E\mathcal{F}\mathbf P_m^H\mathbf o_{IBI}\|^2,\nonumber \\
	\!\!\!\!&=&\!\!\!\! \mathcal {E}[\mathbf E\mathcal{F}\mathbf P_m^H\mathbf o_{IBI}\mathbf o^H_{IBI}\mathbf P_m\mathcal{F}^H\mathbf E^H],\nonumber \\
	\!\!\!\!&=&\!\!\!\! \mathbf E\mathcal{F}\mathbf P_m^H\mathcal {E}[\mathbf o_{IBI}\mathbf o^H_{IBI}]\mathbf P_m\mathcal{F}^H\mathbf E^H,\nonumber \\
	\!\!\!\!&=&\!\!\!\! \mathbf E\mathcal{F}\mathbf P_m^H\boldsymbol\alpha_{IBI}\mathbf P_m\mathcal{F}^H\mathbf E^H,
\end{eqnarray}
where $\boldsymbol\alpha_{IBI} \!\!=\!\! \mathcal {E}[\mathbf o_{IBI}\mathbf o^H_{IBI}]$, now using (\ref{eq:7}), we can determine $\boldsymbol\alpha_{IBI}$ as
\begin{eqnarray}\label{eq:68}
	\boldsymbol\alpha_{IBI}\!\!\!&=& \!\!\!\mathcal {E}\Big[\Big\{ \sum_{l=0}^{L-1} \rho_l \mathbf Z_{l}\mathbf{y}_{B,l}\Big\}\Big\{ \sum_{l=0}^{L-1} \rho_l \mathbf Z_{l}\mathbf{y}_{B,l}\Big\}^H\Big],\nonumber \\
	\!\!\!&=&\!\!\! \mathcal {E}\Big[\sum_{l=0}^{L-1}\rho^2_l\mathbf Z_{l}\mathcal{E}\{\mathbf{y}_{B,l}\mathbf{y}^H_{B,l}\}\mathbf Z^H_{l}\Big],
\end{eqnarray}
Since $\mathbf Z_l$ has a complex Gaussian distribution i.e. $\mathbb{C}\mathcal{N}(0,1)$ and also $\mathbf Z_l$ and $\mathbf{y}_{B,l}$ are uncorrelated, we can write the above equation as follows
\begin{eqnarray}\label{eq:69}
	\boldsymbol\alpha_{IBI}=\sum_{l=0}^{L-1}\rho^2_l\mathcal{E}\{\mathbf{y}_{B,l}\mathbf{y}^H_{B,l}\},
\end{eqnarray}
$\mathcal {E}\{\mathbf{y}_{B,l} \mathbf{y}^H_{B,l}\}$ is dependent on the signal type of the last bock, where we assume it  is also occupied by an FBMC symbol with the same power, then we have 
\begin{eqnarray}\label{eq:70}
	\mathcal {E}\{\mathbf{y}_{B,l} \mathbf{y}^H_{B,l}\} \!\!\!\!&=&\!\!\!\! \mathcal{E}\|\mathbf {P}_{(l)}\mathbf {b}_{last}\|^2\!\!=\!\!\textrm{Tr}\big[\mathbf {P}_{(l)}\mathcal {E}\{\mathbf {b}_{last}{\mathbf {b}^H_{last}}\}\mathbf {P}^H_{(l)}\big],\nonumber \\
	\!\!\!&=&\!\!\! \delta^2\textrm{Tr}\big[\mathbf {P}_{(l)}\mathbf {P}^H_{(l)}\big]=\delta^2\textrm{Tr}\big[\mathbf P^{corr}_{(l)}],\nonumber\\
	\!\!\!&=&\!\!\!\delta^2 P^{corr}_{(l)},
\end{eqnarray}
where $\mathbf{P}_{(l)} \!\!\!=\!\!\![\mathbf{P}_{(last-l)}; \mathbf 0_{(M+K-1)N-l \times MN}]$ in which $\mathbf{P}_{(last-l)}$ contains the last $l$-th rows of $\mathbf{P}$ also $\mathbf {b}_{last}$  is the symbol (after IDFT) in the last block and that $\mathcal {E}\{\mathbf {b}_{last}{\mathbf {b}^H_{last}}\} = \delta^2 \mathbf I$. Substituting (\ref{eq:70}) in (\ref{eq:69}), we obtain
\begin{eqnarray}\label{eq:71}
	\alpha_{IBI} \!\!\!\!&=&\!\!\!\!\delta^2\sum_{l=0}^{L-1}\rho^2_l P^{corr}_{(l)},%\nonumber \\
	%&=& \delta^2 P^{corr}_{h}
\end{eqnarray}
Since $P^{corr}_{(l)}$ is a scalar value, therefore $\alpha_{IBI}$ is also a scalar value. Substituting it into (\ref{eq:67}), yields
\begin{eqnarray}\label{eq:72}
	\boldsymbol \gamma_{IBI,m}= \alpha_{IBI}\mathbf E\mathcal{F}\mathbf P_m^H\mathbf P_m\mathcal{F}^H\mathbf E^H,
\end{eqnarray}
By taking the $n^{th}$ diagonal element of $\boldsymbol \gamma_{IBI,m}$, we derive the MSE due to IBI as
\begin{eqnarray}\label{eq:73}
	\gamma_{IBI,m,n}\!\!\!\!&=&\!\!\!\!\alpha_{IBI}\|\mathbf E\mathcal{F}\mathbf P_m^H\mathbf P_m\mathcal{F}^H\mathbf E^H\|_n,\nonumber\\
	\!\!\!\!&=&\!\!\!\!\alpha_{IBI}|E_n|^2.
\end{eqnarray}
where $\|\mathcal{F}\mathbf P_m^H\mathbf P_m\mathcal{F}^H\|_n=\|\mathbf I_{N\times N}\|_n$. If we further notice that the elements of $\mathbf{P}_{(last-l)}$ are very small and contains the last $l$ rows of matrix $\mathbf W_{K-1}$. Therefore, $\mathbf P^{corr}_{(l)}$ will be a diagonal matrix with first $l$-th diagonal elements being the square of the last $l$-th elements of filter $\mathbf w$ i.e., $w^2_{KN-l},w^2_{KN-l+1},...,w^2_{KN-1}$. We can therefore represent $\mathbf P^{corr}_{(l)}$ as follows
\begin{eqnarray}\label{eq:74}
	\mathbf P^{corr}_{(l)}\!\! \!&=&\!\!\! \textrm{diag}[w^2_{KN-l},w^2_{KN-l+1},...,w^2_{KN-1},\nonumber \\
	&&\mathbf 0_{(K+M-1)N-l}]
\end{eqnarray}
Using (\ref{eq:74}), we can have the following approximation
\begin{eqnarray}\label{eq:75}
	\gamma_{IBI,m,n} \approx \delta^2 |{E}_{n}|^2\sum_{l=0}^{L-1}\rho_l^2  \sum_{k=0}^{l-1}w_{KN-1-k}^2.
\end{eqnarray}
%\begin{eqnarray}\label{eq:40}
%\boldsymbol\gamma_{IBI,m,n}&=&\mathcal {E}\|\boldsymbol{\psi}_{IBI,m}\|_n^2 = \mathcal {E}\|\mathbf E\mathcal{F}\mathbf P_m^H\mathbf o_{IBI}\|_n^2\nonumber, \\
%&=&\alpha_{IBI}|E_n|^2.
%\end{eqnarray}
%where $\alpha_{IBI} =\delta^2\mathlarger{\sum_{l=0}^{L-1}}\rho^2_l P^{corr}_{(l)}$.\\
%{\bf Proof:} See Appendix \ref{sec:a6}.
%SECTION III-A-VI MSE of Noise
\subsubsection{MSE of Noise}
%The variance of noise is
We can write the variance of the noise from (\ref{eq:35}) and (\ref{eq:27}) as
\begin{eqnarray}\label{eq:76}
	\boldsymbol\gamma_{noise,m}\!\!\!\!&=&\!\!\!\!\mathcal {E}\|\boldsymbol{\psi}_{noise,m}\|^2 = \mathcal {E}\|\mathbf E\mathcal{F}\mathbf P_m^H\mathbf n\|^2,\nonumber \\
	\!\!\!\!&=&\!\!\!\! \mathcal {E}[\mathbf E\mathcal{F}\mathbf P_m^H\mathbf n\mathbf n^H\mathbf P_m\mathcal{F}^H\mathbf E^H],\nonumber \\
	\!\!\!\!&=&\!\!\!\! \sigma^2\mathbf E\mathcal{F}\mathbf P_m^H\mathbf P_m\mathcal{F}^H\mathbf E^H,%\nonumber \\
	%&=&\sigma^2\mathbf E\mathbf E^H
\end{eqnarray}
where $\mathcal{E}\{\mathbf n \mathbf n^H\}=\mathcal{E}\|\mathbf n\|^2=\sigma^2$ since $\mathbf{n}$ is Gaussian noise with each element having zero mean and variance $\sigma^2$. Taking the $n^{th}$ diagonal element of (\ref{eq:76}), we have
\begin{eqnarray}\label{eq:77}
	\boldsymbol\gamma_{noise,m,n}\!\!\!\!&=&\!\!\!\!\sigma^2\|\mathbf E\mathcal{F}\mathbf P_m^H\mathbf P_m\mathcal{F}^H\mathbf E^H\|_n=\sigma^2|E_n|^2.
\end{eqnarray}
where $\|\mathcal{F}\mathbf P_m^H\mathbf P_m\mathcal{F}^H\|_n=\|\mathbf I_{N\times N}\|_n$.
%\begin{eqnarray}\label{eq:41}
%\boldsymbol\gamma_{noise,m,n}&=&\mathcal {E}\|\boldsymbol{\psi}_{noise,m}\|_n^2 = \mathcal {E}\|\mathbf E\mathcal{F}\mathbf P_m^H\mathbf n\|_n^2,\nonumber \\
%&=&\sigma^2|E_n|^2.
%\end{eqnarray}
%{\bf Proof:} See Appendix \ref{sec:a2}.
%%SECTION III-B Interference Power in case of inverse filter
\subsection{Interference / noise power in case of inverse filter}\label{sec3}
As can be seen from (\ref{eq:33}) that with the inverse filter, the estimated symbol is accompanied with MMSE estimation bias, filter distortion and IBI due to channel multipath effect and noise i.e.,
\begin{eqnarray}\label{eq:42}
\mathbf {\hat s}_m=\underbrace{\mathbf s_m}_{\textrm{Desired Signal}}+\underbrace{\boldsymbol{\psi}_{resd,m}}_{\textrm{MMSE Estimation Bias}}\nonumber \\
+\!\!\!\!\!\!\!  \underbrace{\boldsymbol{\psi}_{fd,m}}_{\textrm{Filter Distortion by Multipath}}\!\!\!+ \underbrace{\boldsymbol{\psi}_{IBI,m}}_{\textrm{IBI by Multipath}}+ \underbrace{\boldsymbol{\psi}_{noise,m}}_{\textrm{Noise}}.
\end{eqnarray}
\indent Similar to (\ref{eq:35}), we can write the MSE of the $n$-th modulation symbol estimation in the $m$-th FBMC/QAM symbol as follows
\begin{eqnarray}\label{eq:43}
\gamma_{tot,m,n} &=& \mathcal {E}||{\hat s}_{m,n}- {s}_{m,n}||^2 \nonumber \\
&=& \mathcal {E}\big[ \|\boldsymbol{\psi}_{resd,m}\|_n^2+ \|\boldsymbol{\psi}_{fd,m}\|_n^2+\|\boldsymbol{\psi}_{IBI,m}\|_n^2\nonumber \\
&&+\|\boldsymbol{\psi}_{noise,m}\|_n^2\big].
\end{eqnarray}
%SECTION III-B-I MSE of MMSE Estimation Bias
\subsubsection{MSE of signal estimation bias}\label{sec3a}
This residual interference caused by the MMSE equalizer is same as (\ref{eq:55}) since it is independent from the effect of inverse filter matrix $\mathbf R$. Hence, the interference power of the MMSE estimation biased is
\begin{eqnarray}\label{eq:44}
\gamma_{resd,m,n}\!\!\!\!&=&\!\!\!\!\mathcal {E}\|\boldsymbol{\psi}_{resd,m}\|_n^2=\delta^2\bigg[\frac{\nu^2\sigma^4}{(\delta^2|C_n|^2 + \nu\sigma^2)^2}\bigg].
\end{eqnarray}
%SECTION III-B-II MSE of filter distortion
\subsubsection{MSE of Filter Distortion due to multipath channel}\label{sec3c}
%The variance of interference caused by filter distortion due to channel multipath effect is
From (\ref{eq:43}) and (\ref{eq:33}), the variance of interference caused by filter distortion due to channel multipath effect in case of inverse filter is as
\begin{eqnarray}\label{eq:78}
	\boldsymbol\gamma_{fd,m}\!\!\!\!&=&\!\!\!\!\mathcal {E}\|\boldsymbol{\psi}_{fd,m}\|^2 = \mathcal {E}\|\mathbf E\mathcal{F}\mathbf R_m\mathbf P_m^H \mathbf o_{fd}\|^2,\nonumber \\
	\!\!\!\!&=&\!\!\!\! \mathcal {E}[\mathbf E\mathcal{F}\mathbf R_m\mathbf P_m^H \mathbf o_{fd}\mathbf o^H_{fd}\mathbf P_m\mathbf R_m^H\mathcal{F}^H\mathbf E^H],\nonumber \\
	\!\!\!\!&=&\!\!\!\! \mathbf E\mathcal{F}\mathbf R_m\mathbf P_m^H \mathcal{E}[\mathbf o_{fd}\mathbf o^H_{fd}]\mathbf P_m\mathbf R_m^H\mathcal{F}^H\mathbf E^H,\nonumber \\
	\!\!\!\!&=&\!\!\!\! \mathbf E\mathcal{F}\mathbf R_m\mathbf P_m^H \boldsymbol \alpha_{fd}\mathbf P_m\mathbf R_m^H\mathcal{F}^H\mathbf E^H,
\end{eqnarray}
From (\ref{eq:64}), we know %$\boldsymbol\alpha_{fd}=\mathcal{E}[\mathbf o_{fd}\mathbf o^H_{fd}]$ is as follows:
\begin{eqnarray}\label{eq:79}
	\boldsymbol\alpha_{fd}\!=\!\mathcal{E}[\mathbf o_{fd}\mathbf o^H_{fd}]\!=\! \delta^2\sum_{l=0}^{L-1}\rho^2_lT^{\downarrow l},
\end{eqnarray}
where $T^{\downarrow l}\!\!=\!\! \textrm{Tr}[\Delta\mathbf{P}^{\downarrow l}\Delta\mathbf{P}^{\downarrow lH}]$. Since $T^{\downarrow l}$ is a scalar value, therefore $\boldsymbol\alpha_{fd}$ is also a scalar value. Now substituting (\ref{eq:79}) into (\ref{eq:78}), yields
\begin{eqnarray}\label{eq:80}
	\boldsymbol \gamma_{fd,m}= \alpha_{fd}\mathbf E\mathcal{F}\mathbf R_m\mathbf P_m^H\mathbf P_m\mathbf R_m^H\mathcal{F}^H\mathbf E^H,
\end{eqnarray}
By taking the $n^{th}$ diagonal element of $\boldsymbol \gamma_{fd,m}$, we obtain
\begin{eqnarray}\label{eq:81}
	\gamma_{fd,m,n}\!\!\!\!&=&\!\!\!\!\alpha_{fd}\|\mathbf E\mathcal{F}\mathbf R_m\mathbf P_m^H\mathbf P_m\mathbf R_m^H\mathcal{F}^H\mathbf E^H\|_n,\nonumber \\
	\!\!\!\!&=&\!\!\!\!\alpha_{fd}|E_n|^2\zeta_{m,n}.
\end{eqnarray}
where $\|\mathcal{F}\mathbf R_m\mathbf P_m^H\mathbf P_m\mathbf R_m^H\mathcal{F}^H\|_n=\zeta_{m,n}\|\mathbf I_{N\times N}\|_n$. 
%\begin{eqnarray}\label{eq:45}
%\boldsymbol\gamma_{fd,m,n}&=&\mathcal {E}\|\boldsymbol{\psi}_{fd,m}\|_n^2 = \mathcal {E}\|\mathbf E\mathcal{F}\mathbf R_m\mathbf P_m^H \mathbf o_{fd}\|_n^2,\nonumber \\
%&=&\alpha_{fd}|E_n|^2\zeta_{m,n}.
%\end{eqnarray}
%where $\alpha_{fd}=\delta^2 \mathlarger{\sum_{l=0}^{L-1}}\rho^2_l\textrm{Tr}\{\Delta\mathbf{P}^{\downarrow l}(\Delta\mathbf{P}^{\downarrow l})^H\}$.\\
%{\bf Proof:} See Appendix \ref{sec:a8}.\\
%The enhancement factor can be determined as
%\begin{eqnarray}\label{eq:46}
%\zeta_{m,n}=\|\mathcal{F}\mathbf R_m\mathbf P_m^H\mathbf P_m\mathbf R_m^H\mathcal{F}^H\|_n.
%\end{eqnarray}
%SECTION III-B-III MSE of IBI
\subsubsection{MSE of IBI}\label{sec3d}
Let us consider the case when we have inter-block interference due to the lack of guard time. %The variance of interference caused by IBI in this case is
From (\ref{eq:43}) and (\ref{eq:33}), we can write the variance of the interference caused by IBI in case of inverse filter as
\begin{eqnarray}\label{eq:82}
\boldsymbol\gamma_{IBI,m}\!\!\!\!&=&\!\!\!\!\mathcal {E}\|\boldsymbol{\psi}_{IBI,m}\|^2 
\!=\! \mathcal {E}\|\mathbf E\mathcal{F}\mathbf R_m\mathbf P_m^H\mathbf o_{IBI}\|^2,\nonumber \\
\!\!\!\!&=&\!\!\!\! \mathcal {E}[\mathbf E\mathcal{F}\mathbf R_m\mathbf P_m^H\mathbf o_{IBI}\mathbf o^H_{IBI}\mathbf P_m\mathbf R_m^H\mathcal{F}^H\mathbf E^H],\nonumber \\
\!\!\!\!&=&\!\!\!\! \mathbf E\mathcal{F}\mathbf R_m\mathbf P_m^H\mathcal {E}[\mathbf o_{IBI}\mathbf o^H_{IBI}]\mathbf P_m\mathbf R_m^H\mathcal{F}^H\mathbf E^H,\nonumber\\
\!\!\!\!&=&\!\!\!\! \mathbf E\mathcal{F}\mathbf R_m\mathbf P_m^H\boldsymbol\alpha_{IBI}\mathbf P_m\mathbf R_m^H\mathcal{F}^H\mathbf E^H,
\end{eqnarray}
From (\ref{eq:71}), we already know  %$\boldsymbol\alpha_{IBI}$ is as follows
\begin{eqnarray}\label{eq:83}
\alpha_{IBI} \!\!\!&=& \!\!\!\delta^2\sum_{l=0}^{L-1}\rho_l^2 P^{corr}_{(l)},
\end{eqnarray}
Since $P^{corr}_{(l)}$ is a scalar value, therefore $\alpha_{IBI}$ is also a scalar value. Substituting (\ref{eq:83}) into (\ref{eq:82}), yields
\begin{eqnarray}\label{eq:84}
\boldsymbol \gamma_{IBI,m}= \alpha_{IBI}\mathbf E\mathcal{F}\mathbf R_m\mathbf P_m^H\mathbf P_m\mathbf R_m^H\mathcal{F}^H\mathbf E^H,
\end{eqnarray}
By taking the $n^{th}$ diagonal element of $\boldsymbol \gamma_{IBI,m}$, we derive the MSE of IBI as
\begin{eqnarray}\label{eq:85}
\gamma_{IBI,m,n}\!\!\!&=&\!\!\!\alpha_{IBI}\|\mathbf E\mathcal{F}\mathbf R_m\mathbf P_m^H\mathbf P_m\mathbf R_m^H\mathcal{F}^H\mathbf E^H\|_n,\nonumber\\
\!\!\!&=&\!\!\!\alpha_{IBI}|E_n|^2\zeta_{m,n}.
\end{eqnarray}
where $\|\mathcal{F}\mathbf R_m\mathbf P_m^H\mathbf P_m\mathbf R_m^H\mathcal{F}^H\|_n=\zeta_{m,n}\|\mathbf I_{N\times N}\|_n$.
%\begin{eqnarray}\label{eq:47}
%\boldsymbol\gamma_{IBI,m,n}\!\!\!&=&\!\!\!\mathcal {E}\|\boldsymbol{\psi}_{IBI,m}\|_n^2 = \mathcal {E}\|\mathbf E\mathcal{F}\mathbf R_m\mathbf P_m^H\mathbf o_{IBI}\|_n^2,\nonumber \\
%\!\!\!&=&\!\!\!\alpha_{IBI}|E_n|^2\zeta_{m,n}.
%\end{eqnarray}
%where $\alpha_{IBI} =\delta^2\mathlarger{\sum_{l=0}^{L-1}}\rho^2_l P^{corr}_{(l)}$.\\
%{\bf Proof:} See Appendix \ref{sec:a9}.
%SECTION III-B-IV MSE of NOISE
\subsubsection{MSE of Noise}\label{sec3b}
\begin{figure}[h]
	\centering
	\begin{subfigure}[b]{0.235\textwidth}
		\includegraphics[width=\textwidth]{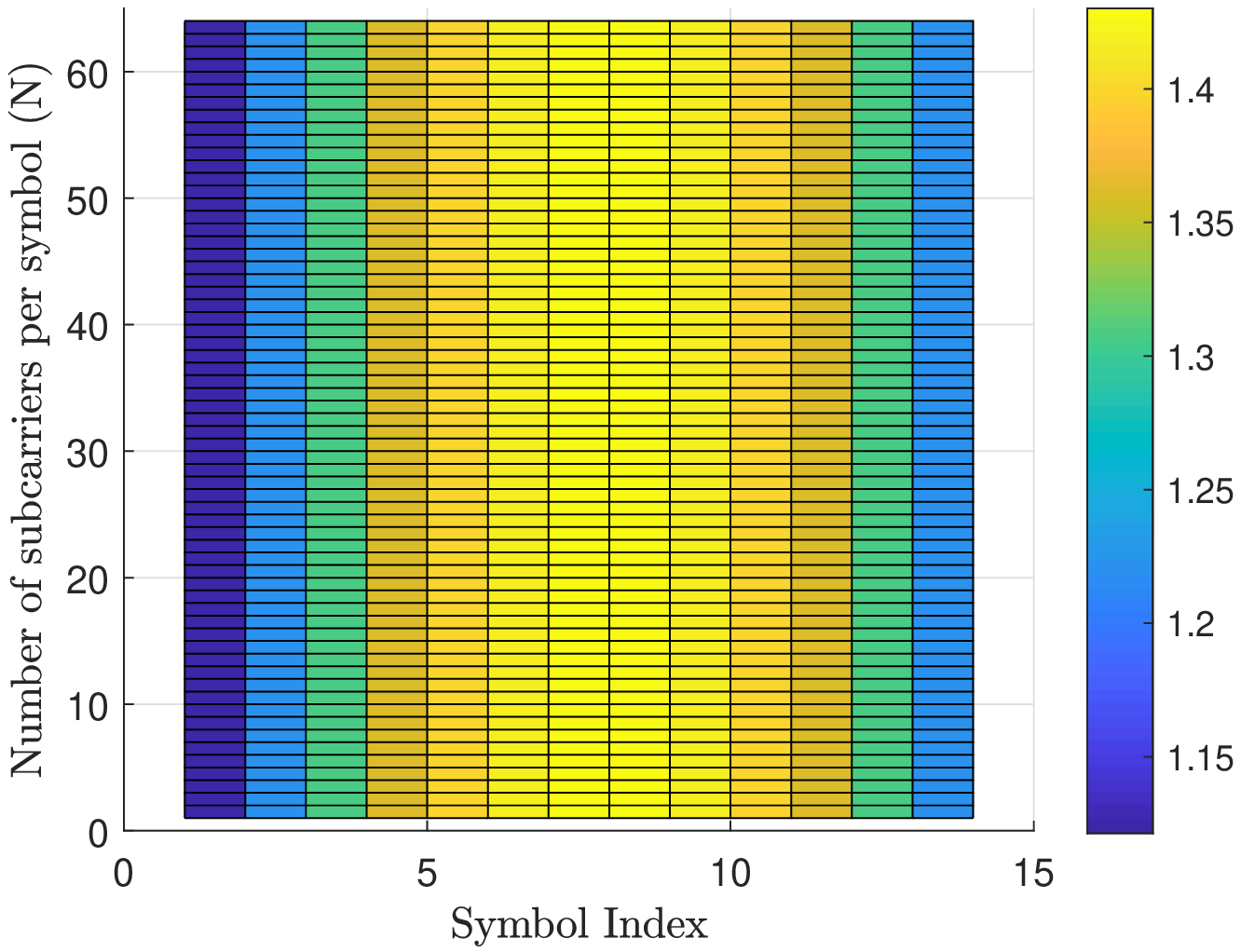}
		\caption{$\zeta_{m,n}$$\,$/$\,$subcarrier$\,$/$\,$symbol}
		\label{fig4a}
	\end{subfigure}
	~ %add desired spacing between images, e. g. ~, \quad, \qquad, \hfill etc. 
	\begin{subfigure}[b]{0.235\textwidth}
		\includegraphics[width=\textwidth]{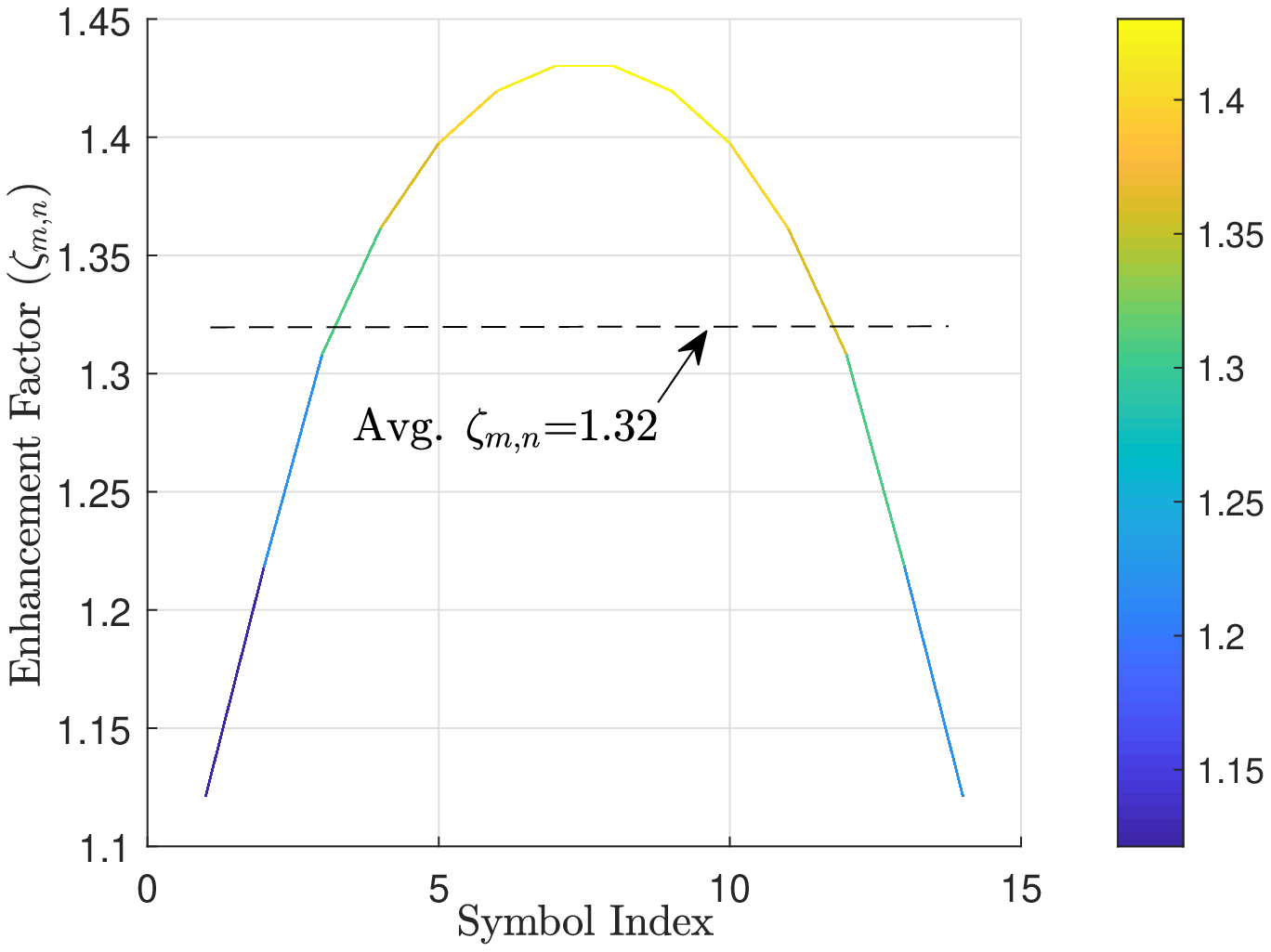}
		\caption{$\zeta_{m,n}$$\,$/$\,$symbol}
		\label{fig4b}
	\end{subfigure}
	\caption{Noise/Interference enhancement factor $\zeta_{m,n}$ for $N\!\!=\!\!64$ and $M\!\!=\!\!14$}\label{fig4}
\end{figure}

%The variance of noise in case of inverse filter is
From (\ref{eq:43}) and (\ref{eq:33}), the variance of noise in case of inverse filter as
\begin{eqnarray}\label{eq:86}
\boldsymbol\gamma_{noise,m}\!\!\!&=&\!\!\!\mathcal {E}\|\boldsymbol{\psi}_{noise,m}\|^2 = \mathcal {E}\|\mathbf E\mathcal{F}\mathbf R_m\mathbf P_m^H\mathbf n\|^2,\nonumber \\
\!\!\!&=&\!\!\! \mathcal {E}[\mathbf E\mathcal{F}\mathbf R_m\mathbf P_m^H\mathbf n\mathbf n^H\mathbf P_m\mathbf R_m^H\mathcal{F}^H\mathbf E^H],
\end{eqnarray}
As $\mathcal{E}\{\mathbf n \mathbf n^H\}=\mathcal{E}\|\mathbf n\|^2=\sigma^2$ since $\mathbf{n}$ is Gaussian noise with each element having zero mean and variance $\sigma^2$. Taking the $n^{th}$ diagonal element of (\ref{eq:86}), we obtain
\begin{eqnarray}\label{eq:87}
\boldsymbol\gamma_{noise,m,n}\!\!\!\!&=&\!\!\!\!\sigma^2\|\mathbf E\mathcal{F}\mathbf R_m\mathbf P_m^H\mathbf P_m\mathbf R_m^H\mathcal{F}^H\mathbf E^H\|_n,\nonumber\\
\!\!\!\!&=&\!\!\!\!\sigma^2|E_n|^2\zeta_{m,n}.
\end{eqnarray}
where $\|\mathcal{F}\mathbf R_m\mathbf P_m^H\mathbf P_m\mathbf R_m^H\mathcal{F}^H\|_n=\zeta_{m,n}\|\mathbf I_{N\times N}\|_n$.\\
%\begin{eqnarray}\label{eq:48}
%\boldsymbol\gamma_{noise,m,n}&=&\mathcal {E}\|\boldsymbol{\psi}_{noise,m}\|_n^2= \mathcal {E}\|\mathbf E\mathcal{F}\mathbf R_m\mathbf P_m^H\mathbf n\|_n^2,\nonumber \\
%&=&\sigma^2|E_n|^2\zeta_{m,n}.
%\end{eqnarray}
%%MSE of distorted filter
%{\bf Proof:} See Appendix \ref{sec:a7}.\\
\indent Note that the $\zeta_{m,n}$ is the noise / interference enhancement factor which is introduced when we use an inverse filter matrix at the receiver. The noise/interference enhancement factor per subcarrier per symbol ($\zeta_{m,n}$$\,$/$\,$subcarrier$\,$/$\,$symbol) %$(\textrm{we have assumed } N=64 \textrm{ and } M=14)$
is presented in Fig. \ref{fig4}.
It can be seen that the noise enhancement factor is constant for every subcarrier in each symbol and its effect is maximum for the symbols in the middle of the FBMC/QAM data block as shown in Fig. \ref{fig4a}. However, the impact is not significant since the average enhancement factor in a block is 1.32 as can be seen from Fig. \ref{fig4b}.
%SECTION IV Complexity Analysis
\section{Complexity Analysis}\label{comp_ana}
In this section we have presented the complexity analysis of the FBMC/QAM system with and without inverse filter at the receiver. The objective is to determine if there is a significant increase in the complexity of the system with the introduction of the inverse filter matrix at the receiver. Typically the complexity of a system is measured by the number of floating point operations (FLOPS), we however, only focus on the number of real multiplications in our complexity analysis. Since we have already presented the FBMC/QAM transmitter and receiver processes in matrix multiplication form in Section \ref{sec2}, we have adopted the $\textit{naive matrix multiplication algorithm}$ \cite{yuster2005fast} to perform the complexity analysis.
%SECTION IV-A CA for no inverse filter
\subsection {Complexity analysis in case of no inverse filter}
To determine the complexity of the FBMC/QAM system without inverse filter, we have to look at the structure of the system as shown in Fig. \ref{fig1}.
%SECTION IV-A-I Transmitter Complexity]
\subsubsection{Transmitter Complexity}
Since our model follows a block based processing approach, the input to the system is a vector of size $MN$. Each QAM symbol in the FBMC/QAM transmitter requires an $N$-point IDFT operation. The most efficient FFT algorithm i.e. split-radix requires $Nlog_2N-3N+4$ real multiplications \cite{Gerzaguet2017,8067703}. The complex vector $\mathbf b$ at the output of the IDFT processor is then processed through the transmit filter matrix $\mathbf P$. Since, the structure of $\mathbf P$ is sparse as shown in (\ref{eq:2}), the number of real multiplications involved in filtering operation is determined as $\sum_{k=1}^{MN}\mathbf P_k \mathbf b_k = 2MNK$ per block. Where $\mathbf P_k$ is the number of nonzero elements in the $k^{th}$ column of matrix $\mathbf P$, and $\mathbf b_k$ is the number of nonzero elements in the $k^{th}$ row of vector $\mathbf b$. The total number of real multiplications involved in the transmitter per complex-valued symbol is
\begin{eqnarray}\label{eq:49}
\mathbf C_{\textrm{Tx}} = Nlog_2N+(2K-3)N+4.
\end{eqnarray}
\begin{figure}[h]
	\centering
	\includegraphics[scale=0.6]{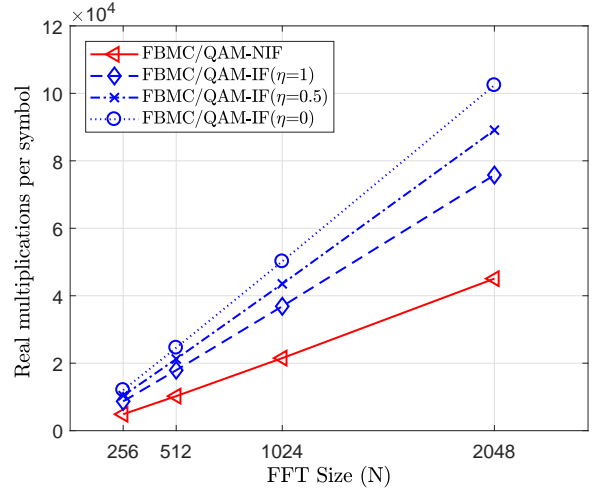}
	\caption{Complexity comparison of FBMC/QAM with and without inverse filter for $K\!\!=\!\!5$ and $M\!\!=\!\!14$}
	\label{fig6}
\end{figure}
\begin{figure*}[t]
	\centering
	\begin{subfigure}[b]{0.3\textwidth}
		\includegraphics[width=\textwidth]{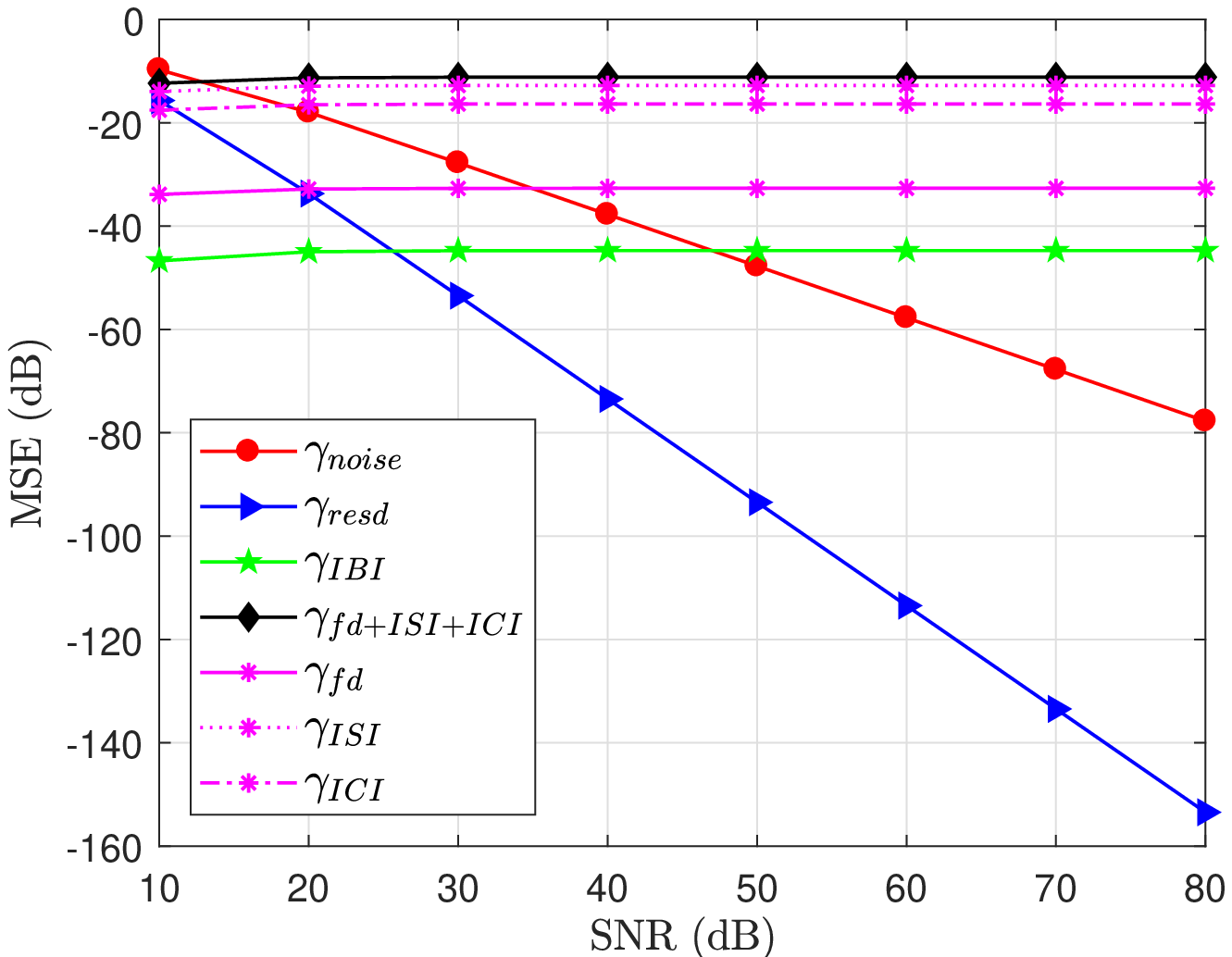}
		\caption{Individual MSE}
		\label{fig5a}
	\end{subfigure}
	~ %add desired spacing between images, e. g. ~, \quad, \qquad, \hfill etc. 
	\begin{subfigure}[b]{0.3\textwidth}
		\includegraphics[width=\textwidth]{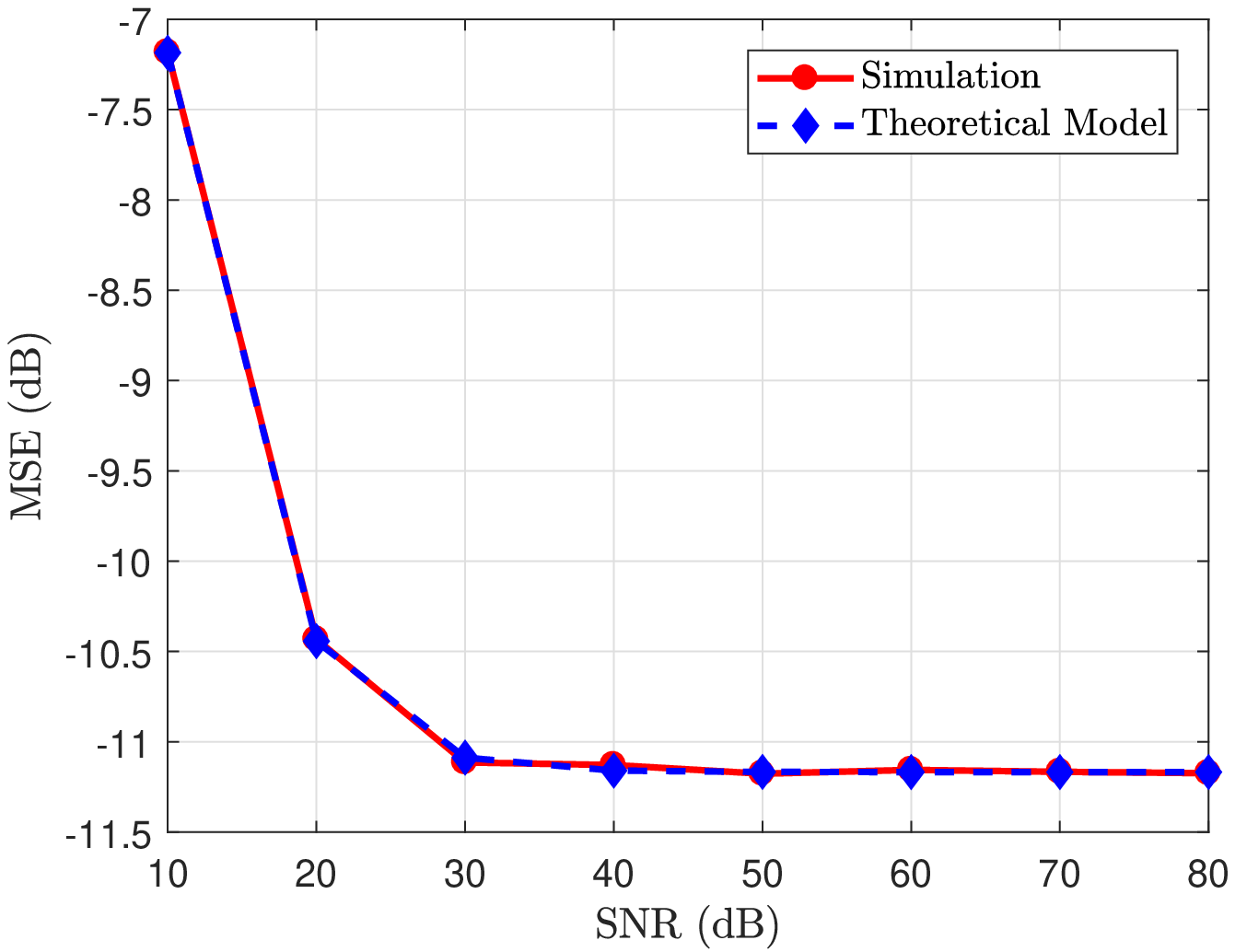}
		\caption{Composite MSE}
		\label{fig5b}
	\end{subfigure}
	~ %add desired spacing between images, e. g. ~, \quad, \qquad, \hfill etc. 
	%	\begin{subfigure}[b]{0.2\textwidth}
	%		\includegraphics[width=\textwidth]{no_inv_filt_const.pdf}
	%		\caption{Tx and Rx Constellation}
	%		\label{fig:const_no_invF}
	%	\end{subfigure}
	%	~ %add desired spacing between images, e. g. ~, \quad, \qquad, \hfill etc. 
	%	%(or a blank line to force the subfigure onto a new line)
	\begin{subfigure}[b]{0.3\textwidth}
		\includegraphics[width=\textwidth]{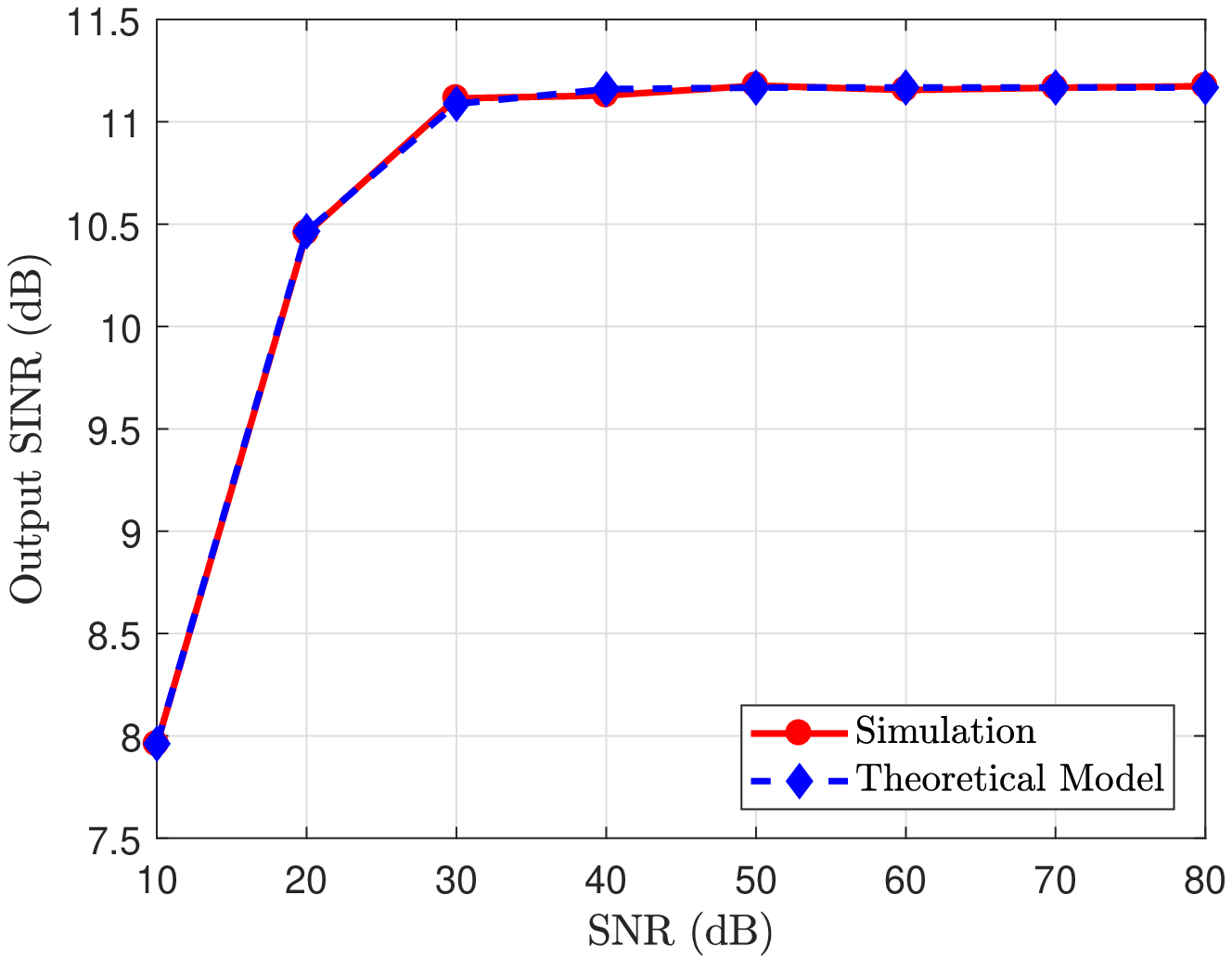}
		\caption{SNR Vs SINR}
		\label{fig5c}
	\end{subfigure}
	%\caption{Constellations and SNR vs SINR}\label{fig:const_snr_sinr_no_invF}
	\caption{Performance of FBMC/QAM without Inverse Filter}\label{fig5}
\end{figure*}
\begin{figure*}[t]
	\centering
	\begin{subfigure}[b]{0.3\textwidth}
		\includegraphics[width=\textwidth]{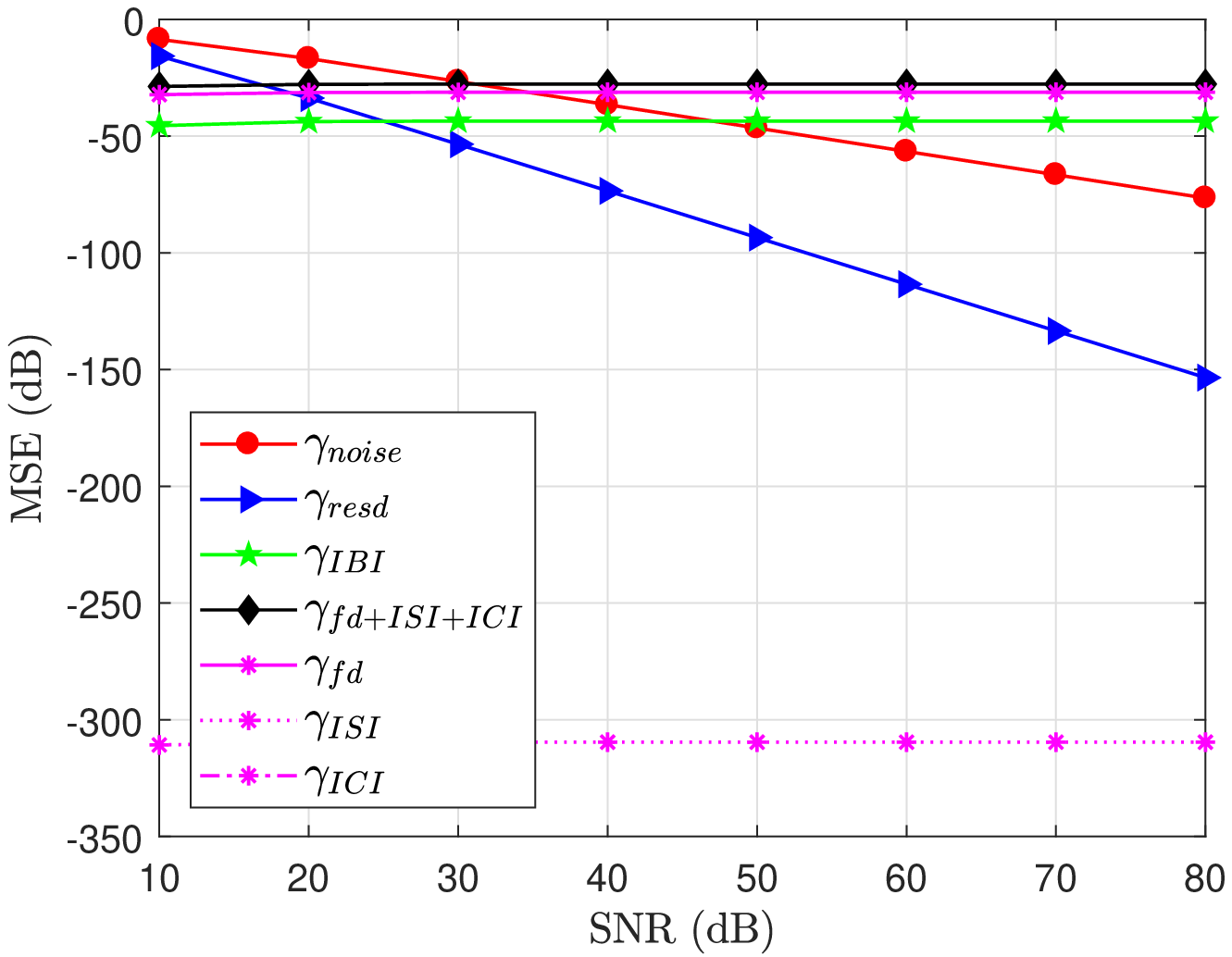}
		\caption{Individual MSE Components}
		\label{fig7a}
	\end{subfigure}
	~ %add desired spacing between images, e. g. ~, \quad, \qquad, \hfill etc. 
	\begin{subfigure}[b]{0.3\textwidth}
		\includegraphics[width=\textwidth]{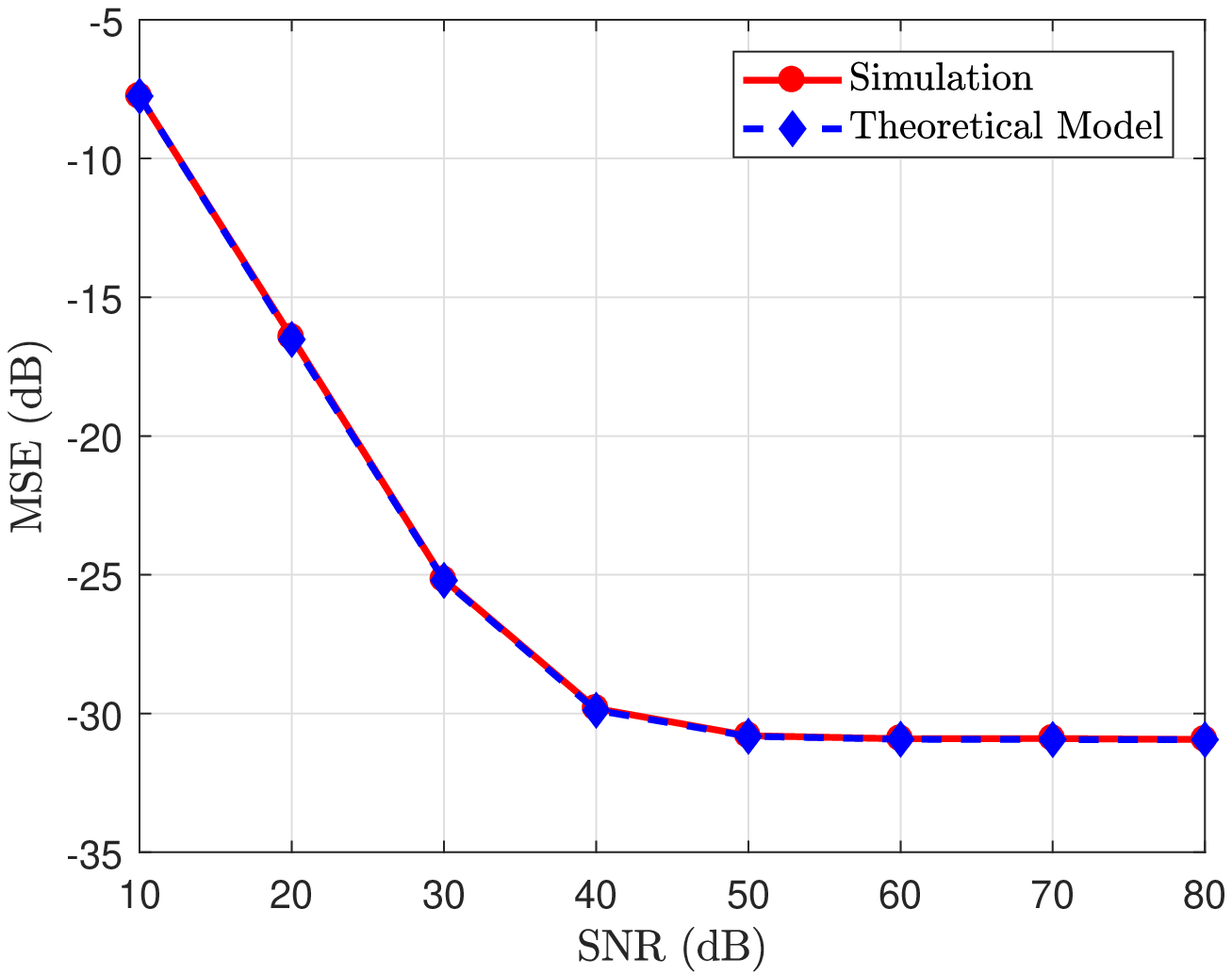}
		\caption{Composite MSE}
		\label{fig7b}
	\end{subfigure}
	~ %add desired spacing between images, e. g. ~, \quad, \qquad, \hfill etc. 
	%	\begin{subfigure}[b]{0.4\textwidth}
	%		\includegraphics[width=\textwidth]{inv_filt_const.pdf}
	%		\caption{Tx and Rx Constellation}
	%		\label{fig:const_invF}
	%	\end{subfigure}
	%	~ %add desired spacing between images, e. g. ~, \quad, \qquad, \hfill etc. 
	%	%(or a blank line to force the subfigure onto a new line)
	\begin{subfigure}[b]{0.3\textwidth}
		\includegraphics[width=\textwidth]{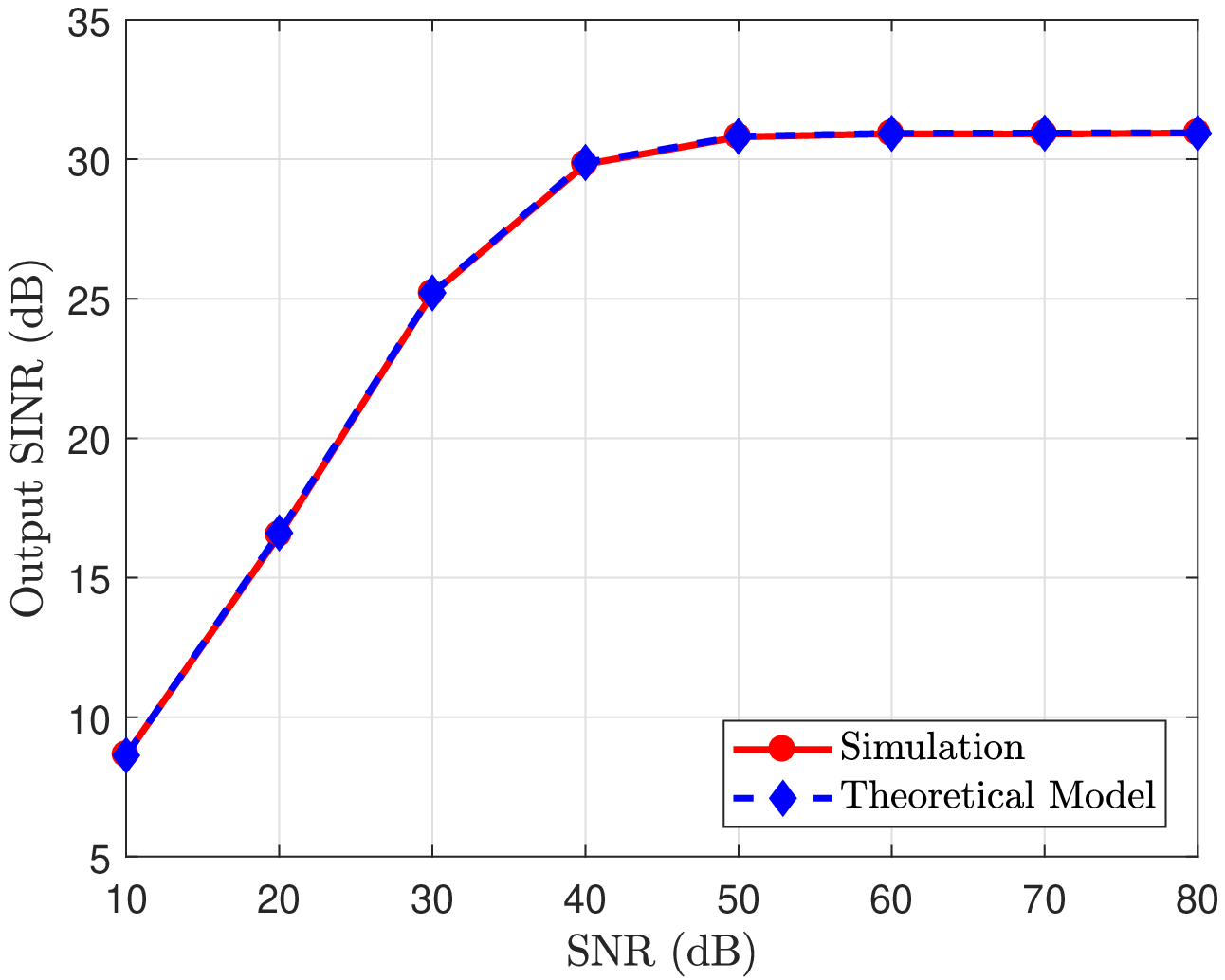}
		\caption{SNR Vs SINR}
		\label{fig7c}
	\end{subfigure}
	\caption{Performance of FBMC/QAM with Inverse Filter}\label{fig7}
\end{figure*}
%SECTION IV-A-II Receiver Complexity
\subsubsection{Receiver Complexity}
It can be seen from Fig. \ref{fig1}, the transmitted signal $\mathbf o$, after passing through the channel, is received by the receiver as a complex vector $\mathbf r$ and is processed by the receiver filter. Using the $\textit{naive matrix multiplication algorithm}$, the number of real multiplications involved in this stage is $2MNK$ per block. After serial to parallel conversion, each symbol is processed by a $N$ point DFT operation resulting in $Nlog_2N-3N+4$ real multiplications for processing one symbol. The complex-valued symbols after the DFT processing are then equalized using $\mathbf E$ as defined in (\ref{eq:24}). The equalization process requires $4MN$ real multiplications to estimate one transmitted FBMC/QAM block. Hence, the total number of real multiplications per complex symbol in the case of no inverse filter ($\textit{NIF}$) at the receiver is
\begin{eqnarray}\label{eq:50}
\mathbf C_{\textrm{Rx}}^{NIF} = Nlog_2N+(2K+1)N+4.
\end{eqnarray}
%SECTION IV-B CA for inverse filter
\subsection{Complexity analysis in case of inverse filter}
The transmitter complexity in this case is the same as (\ref{eq:49}) since inverse filter only increases the complexity of the receiver. %For receiver complexity, 
%\subsubsection{Receiver Complexity}
%it can be seen from Fig. \ref{fig2} that the received signal after the receiver filter matrix is processed using the inverse filter matrix $\mathbf R \in \mathbb{R}^{MN \times MN}$. 
The total number of real multiplications per symbol in case of inverse filter ($\textit{IF}$) at the receiver is
\begin{eqnarray}\label{eq:51}
\mathbf C_{\textrm{Rx}}^{IF} = \mathbf C_{\textrm{Rx}}^{NIF} + \mathbf C_{R}.
\end{eqnarray}
where $\mathbf C_{R}=2MN-\eta N(M-1)$ is the number of additional real multiplications per symbol introduced by the inverse filter and depends on the value of $\eta$ as defined in Sec. \ref{sp_mtx}. %Each symbol after the inverse filter is then processed through a $N$ point DFT operation and then equalized to estimate the transmitted FBMC/QAM block. Hence, 
%\textcolor{red}{The complexity of the receiver in case of inverse filter with different cases of $\eta$ are summarized in Table \ref{table1} as follows}
%\begin{table}[h]
%	\centering
%	\caption{\textcolor{red}{Receiver Complexity with Inverse Filter Matrix $(\mathbf R)$}}
%	\label{table1}
%	\begin{tabular}{|c|c|c|c|}
%		\hline
%		\textbf{$S/No$} & \textbf{$\eta$} & \textbf{$\mathbf C_R$} & \textbf{$\mathbf C_{\textrm{Rx}}^{IF}$}              \\ \hline
%		1             & 0          & $2MN$        & $Nlog_2N+(2K+2M+1)N+4$     \\ \hline
%		2             & 0.5        & $\frac{(3M+1)N}{2}$  & $Nlog_2N+(2K+\frac{3M}{2}+\frac{3}{2})N+4$ \\ \hline
%		3             & 1          & $(M+1)N$     & $Nlog_2N+(2K+M+2)N+4$      \\ \hline
%	\end{tabular}
%\end{table}
It can be seen from (\ref{eq:51}) that the complexity of the receiver with inverse filter ($\mathbf C_{\textrm{Rx}}^{IF}$) depends on the block size ($M$). Hence, the additional complexity will be higher for large block size. %Now considering the prototype filter length $K=5$ and symbols per block $M$ = 14, 
The complexity in terms of real multiplications per symbol with and without the inverse filter is presented in Fig. \ref{fig6}. It is worth mentioning that the worst case or the upper bound of the receiver complexity i.e., Big--$\mathcal O$ is $\mathcal O(Nlog_2N)$ \footnote{Although matrix inversion has a general complexity of $\mathcal O(N^3)$ which can significantly increase the complexity of the receiver. However, since the coefficients in the autocorrelation matrix $\mathbf G$ are constant, we can calculate the inverse filter matrix $\mathbf R= \mathbf G^{-1}\in \mathbb{R}^{MN\times MN}$ off-line.} for both cases and can be determined by dropping the lower order terms and the constant multipliers in (\ref{eq:50}) and (\ref{eq:51}). 
%The upper bound of receiver complexity is $\mathcal O(Nlog_2N)$ with and without the inverse filter.\\
%{\bf Proof:} See Appendix \ref{sec:a10}.
%SECTION V Simulation Results
\section{Simulation Results}
In this section we present the simulation results for MSE and output SINR of the FBMC/QAM system with and without inverse filter along with the BER performance in case of synchronous and asynchronous multi-service scenarios.
%SECTION V-A MSE and output SINR
\subsection{MSE and output SINR}
The performance in terms of MSE and output SINR in a FBMC/QAM system without and with the inverse filter at the receiver can be observed from Fig. \ref{fig5} and Fig. \ref{fig7} respectively. The individual MSE sources like noise, residue from the equalization, IBI, ISI, ICI and filter distortion in the two cases are shown in Fig. \ref{fig5a} and Fig. {\ref{fig7a}} respectively. It can be seen that without inverse filter, the contribution of ICI and ISI (intrinsic interference) is quite significant i.e. around -16.5dB and -13dB respectively.  However, with the use of inverse filter the intrinsic interference becomes negligible i.e., ISI is around -310dB and ICI cannot be even displayed on the same scale.\\
\indent The interference caused by the multipath effect includes filter distortion and IBI which contributes around -32.5dB and -45dB respectively in case of no inverse filter. However, these values increase to around -31.5dB and -43.5dB respectively with the use of inverse filter. The increase in interference is due to the enhancement factor $\zeta_{m,n}$ introduced by the use of inverse filter matrix at the receiver as discussed in Sec. \ref{sec3}.\\
\indent The overall MSE performance of FBMC/QAM system without and with the inverse filter are shown in Fig. {\ref{fig5b}} and Fig. {\ref{fig7b}} respectively. It can be observed that without the inverse filter, the system becomes interference limited beyond SNR=30dB and the MSE is around -11.2dB. However, the use of inverse filter improves the system performance and these values become 50dB and -31dB respectively. The output SINR of the system also improves with the use of inverse filter as can be seen from the Fig. \ref{fig5c} and Fig. \ref{fig7c}. It can also be confirmed that the interference terms in the system model give in (\ref{eq:27}) and (\ref{eq:33}) completely matches with the simulation results, which verifies the accuracy of the derived analytical model.
\begin{figure}[h]
	\begin{center}
		\includegraphics[scale=0.6]{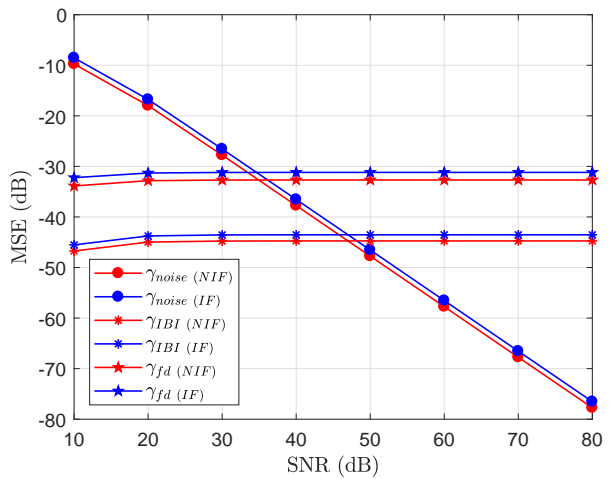}
		\caption{Noise \& Interference Enhancement}\label{fig8}
	\end{center}
\end{figure}
\\\indent The noise and interference ($\gamma_{IBI}$ and $\gamma_{fd}$) enhancement due to the use of inverse filter (\textit{IF}) is illustrated in Fig. \ref{fig8}. It can be seen that the enhancement is very small compared to the no inverse filter case (\textit{NIF}). The interference enhancement is therefore negligible in comparison to the level of performance improvement achieved with the use of inverse filter.
%SECTION V-B BER Performance
\subsection{BER Performance}\label{ber}
\begin{figure}[h]
\begin{center}
		\includegraphics[scale=0.6]{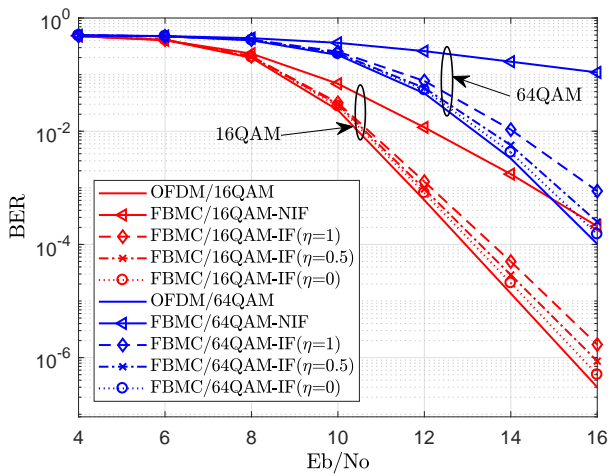}
		\caption{BER performance of OFDM and FBMC/QAM system with synchronous sub-bands}\label{fig9}
\end{center}
\end{figure}

\begin{figure}[th]
	\centering
	\includegraphics[scale=0.55]{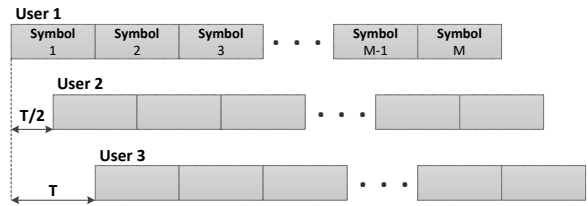}
	\caption{Asynchronous user streams}
	\label{fig10}
\end{figure}

\begin{figure}[h]
	\begin{center}
		\includegraphics[scale=0.6]{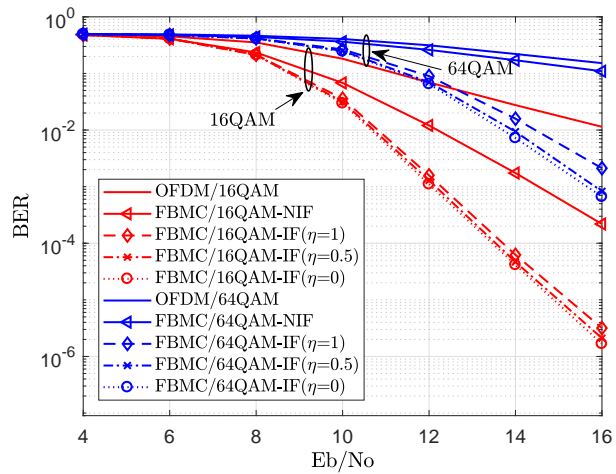}
		\caption{BER performance of OFDM and FBMC/QAM system with asynchronous sub-bands}\label{fig11}
	\end{center}
\end{figure}

The coded results (convolutional code with code rate 1/2 and generator polynomials defined as [133, 171]) for the BER performance of FBMC/QAM system with and without inverse filter are presented for synchronous and asynchronous multi-user transmissions. We have considered a multi-user (multi-service) transmission scenario since next generation wireless systems are expected to provide a flexible framework for heterogeneous services. In such a case, services like mobile broadband (MBB), Internet of things (IoT), ultra reliable communication (URC) may coexist in adjacent sub-bands. To evaluate the performance of FBMC/QAM system with multi-service transmission, we segregate the whole bandwidth into three consecutive sub-bands, each for different user (services). The BER performance of the FBMC/QAM system with and without inverse filtering for synchronous segregated spectrum is shown in Fig. {\ref{fig9}}. We have used conventional OFDM as a baseline scheme to compare the performance of the FBMC/QAM system with and without inverse filter. For a fair comparison between the two systems, the SNR loss, due to the cyclic prefix (overhead) in OFDM, must be considered. For this reason, we have calculated the noise power for both systems as discussed in \cite{azafar}. It can be seen from Fig. {\ref{fig9}} that without inverse filter (FBMC/QAM-NIF), the FBMC/QAM system has poor performance compared to conventional OFDM system due to the presence of intrinsic interference. Since, FBMC/QAM with inverse filter (FBMC/QAM-IF) can cancel the intrinsic interference at the receiver, the system can provide comparable performance to the conventional OFDM system as shown in Fig. {\ref{fig9}}.
\\\indent However, it is likely that the adjacent sub-bands in a multi-service transmission are out of sync since simple IoT devices in future wireless networks may only have coarse synchronization. Thus it is very desirable for the system to be robust against asynchronism between adjacent sub-bands. To evaluate the performance of FBMC/QAM under asynchronous sub-bands, we have considered the timing offset between two adjacent sub-band transmissions to be 50$\%$ of the symbol interval as shown in Fig {\ref{fig10}}. 
Since the second sub-band transmission suffers interference from the first and third sub-band, it is appropriate to investigate the BER performance of the second sub-band user. In case of multi-service asynchronous transmission, the BER performance of FBMC/QAM with and without inverse filter is shown in Fig. {\ref{fig11}}. It can be seen that in case of asynchronous multi-service transmission, FBMC/QAM with inverse filter significantly outperforms the the conventional OFDM system. We can also observe the impact of neglecting the diagonal elements %in the off-diagonal sub-matrices 
in matrix $\mathbf R$ on the system BER performance. It can be seen from Fig. {\ref{fig9}} and Fig. {\ref{fig11}} that for $\eta\!\!=\!\!0$ the BER performance is better than $\eta\!\!=\!\!1$ since we are using all the diagonal elements in the inverse filter matrix. However, the complexity of system is lower for $\eta\!\!=\!\!1$ as neglecting middle $N/2$ elements in the off-diagonal sub-matrices of $\mathbf R$ leads to less additional real multiplications. It is therefore a trade-off between the complexity and the system BER performance i.e., a higher value of $\eta$ leads to lower complexity as well as worse BER performance. Whereas a lower value of $\eta$ leads to higher complexity as well as better BER performance.
In the light of all the results, the improved performance of FBMC/QAM compared to conventional OFDM systems along with its good out-of-band leakage performance, robustness to asynchronous multi-service transmissions and ability to have flexible scheduling on subcarrier level makes it a suitable candidate for next generation wireless applications, especially for massive machine type communications.
%SECTION VI Conclusion
\section{Conclusion}
We have proposed a novel low-complexity interference-free FBMC/QAM system based on matrix inversion of the prototype filters that mitigates the intrinsic interference in a FBMC/QAM system. The proposed system enables the use of complex-valued symbol transmission, while maintaining per subcarrier based filtering. The proposed system is based on a compact matrix model of the FBMC/QAM system, which also laid the ground for an in-depth analysis of the interferences affecting the system when operating in a multipath environment. The interference terms due to channel distortions and the intrinsic behavior of the transceiver model have been derived in detail and analyzed in terms of MSE with and without the inverse filter. It was shown through the theoretical and simulation results that inverse filtering significantly reduces the interference in FBMC/QAM system at the expense of slight enhancement in IBI, interference due to filter distortion caused by multipath channel and noise. The complexity analysis of the system with and without the inverse filter is also provided which shows that complexity in both cases have the same upper bounds. The performance of the system is then evaluated for synchronous and asynchronous multi-service scenarios. Simulation result shows that FBMC/QAM with inverse filter can provide comparable performance to the conventional OFDM system in case of synchronous multi-service transmission while it outperforms OFDM in the asynchronous case. The improved performance of the proposed FBMC/QAM system makes it highly suitable for next generation wireless applications, especially for massive machine type communications.
\bibliographystyle{IEEEtran}
\bibliography{IEEEabrv,varUD}
\end{document}